\documentclass[twocolumn]{aastex631}
\usepackage[T5]{fontenc}
\usepackage{svg}
\usepackage{amsmath}
\usepackage{hyperref}
\newcommand{\corr}[1]{\textcolor{black}{#1}}
\newcommand{\corrtwo}[1]{\textcolor{black}{#1}}

\begin{document}

\title{Effect of molecular hydrogen self-shielding modeling on early Reionization Era galaxies in radiative hydrodynamic cosmological simulations}

\author[0009-0002-2290-8039]{Thịnh Hữu Nguyễn}
\affiliation{Department of Astronomy, University of Illinois at Urbana-
Champaign, Urbana, IL 61801, USA}

\affiliation{Center for AstroPhysical Surveys, National Center for Supercomputing Applications \\
Urbana, IL 61801, USA}

\author[0000-0002-8638-1697]{Kirk S. S. Barrow}
\affiliation{Department of Astronomy, University of Illinois at Urbana-
Champaign, Urbana, IL 61801, USA}

\author[0009-0001-4446-833X]{Susie Byrom}
\affiliation{Department of Astronomy, University of Illinois at Urbana-
Champaign, Urbana, IL 61801, USA}

\author[0009-0009-7224-4462]{Varun Satish}
\affiliation{Department of Astronomy, University of Illinois at Urbana-
Champaign, Urbana, IL 61801, USA}



\begin{abstract}

Accurately modeling molecular hydrogen ($\text{H}_{2}$) is an important task in cosmological simulations because it regulates star formation. One fundamental property of $\text{H}_{2}$ is the ability to self-shield, a phenomenon in which the $\text{H}_{2}$ in the outer layer of a molecular cloud absorbs the photodissociating Lyman-Werner UV radiation and shields the inner $\text{H}_{2}$. Historically, numerical approximations have been utilized to avoid intensive ray-tracing calculations. This paper evaluates the use of the Sobolev-like density-gradient approximation in $\text{H}_{2}$ self-shielding modeling and tests its agreement with a more rigorous adaptive ray-tracing method in cosmological simulations. We ran four high-resolution zoom-in cosmological simulations to investigate the models' effects in the early Reionization Era ($z \geq 12$). We find that the approximation model \corr{returns a higher} $\text{H}_{2}$ photodissociation rate in low gas density environments but \corrtwo{a lower rate} when gas density is high, resulting in low-mass halos having less $\text{H}_{2}$ while high-mass halos having more $\text{H}_{2}$. \corr{The approximation} also hinders star formation in small halos, but it less affects the stellar mass of larger halos. Inside a halo, the discrepancies between the two models regarding $\text{H}_{2}$ fraction, temperature, and stellar mass are radially dependent. On a large scale, the simulations using the approximation have less $\text{H}_{2}$ in the intergalactic medium and \corr{may} experience a slower reionization process. These results show that the Sobolev-like approximation alters properties of galaxies and the large-scale universe \corr{when compared to the ray-tracing treatment}, emphasizing a need for caution when \corr{interpreting results from these two techniques in cosmological simulations}.  

\end{abstract}



\section{Introduction} \label{sec:intro}

%\corr{The third sentence is a stronger starting point. In general, you need to build up to sweeping statements about what is challenging or computationally expensive with citations so if you want to or need to make this point, it should be at the end of a paragraph.}Modeling accurate 3D radiative transfer processes in cosmological simulations has long been a challenging and computationally expensive task. One of the most important processes is the formation and self-shielding nature of molecular hydrogen \corr{Similarly to the first sentence, this sentence makes generalizations about importance without first justifying why this process is more important than others. Throughout this paper, we should built evidence before drawing conclusions. In the introduction, we should be especially careful about using less presumptuous language and use modifiers like "likely" or "it appears that".}. 

%Molecular hydrogen ($\text{H}_{2}$) is an important building block for star formation as well as galaxy evolution and thus demands a rigorous modeling treatment in cosmological simulations. 
Modeling molecular hydrogen ($\text{H}_{2}$) in cosmological simulations demands a rigorous and accurate treatment because it regulates star formation and is an important constituent of the interstellar medium. In the Jeans instability model \citep{Jeans1920}, at a given density, it is easier for a cloud with a lower temperature to collapse because the mass threshold to achieve instability is smaller. Thus, radiative cooling is important for star formation. Depending on the temperature and density, there are multiple radiative cooling mechanisms for a cloud to release its internal energy. For primordial gas clouds with atomic hydrogen (\corr{H}) and helium (\corr{He}) being the dominant component, Bremsstrahlung emission, atomic hydrogen recombination, collisional ionization, and collisional excitation followed by spontaneous de-excitation processes can help cool the gas down to a floor temperature of about $10^{4}$ K, below which the cooling rate is too low \citep{Thoul+1995}. However, if $\text{H}_{2}$ is present, it can further stimulate cooling by opening up new cooling channels through its electronic, rotational, and vibrational energy levels. This allows parts of the gas cloud where $\text{H}_{2}$ is available to cool \corr{down to about 200K \citep{Galli+1998, Glover+2008}} and further contract. 
%leading to the cloud's fragmentation into multiple star-forming sites.
The increased density leads to a decrease in cooling time, which makes the gas collapse faster. This runaway cooling eventually drives the central density up to \corr{$2\times 10^{-4} \text{g}/\text{cm}^{3}$ \citep{Yoshida+2008}}, and a protostar forms. \corr{Previous theoretical studies showed the integral role of $\text{H}_{2}$ as the main cooling mechanism of primordial gas to form early stars in high redshift, metal-free environment \citep{Abel+1997, Glover+2005}. At lower redshift, the correlation between star formation and $\text{H}_{2}$ can be reflected through the Kennicutt-Schmidt (KS) law and observational evidence in nearby disk galaxies \citep{Schmidt+1959, Kennicutt+1998, Wong+2002, Kennicutt+2007}. However, in a more metal-rich environment, metal cooling can also drive the collapse of clouds \citep{Glover+2012, Krumholz+2012}, although it is also important to note that different elements and molecules dominate cooling in different ranges of density and temperature.} 

Ultraviolet (UV) radiation emitted by nearby active stars can photodissociate molecular hydrogen via the Solomon process (Solomon 1965 - private communication reported in \citealt{Field+1966}, \citealt{Stecher+1967}). When absorbing a Lyman-Werner (LW) photon (11.2-13.6 eV), a $\text{H}_{2}$ molecule is excited from the ground state to an excited electronic state. Instead of radiatively decaying back to the ground state, there is about a 15 percent chance that the excited molecule has its electrons decay into the vibrational continuum, which subsequently dissociates the molecule into atomic hydrogen. The dissociating LW flux on molecules can be attenuated through $\text{H}_2$ self-shielding, a phenomenon where the $\text{H}_2$ column density is large enough that there is a dynamic balance between $\text{H}_{2}$ dissociation-recombination in the cloud's outer layer, allowing the $\text{H}_{2}$ in the outer layer to absorb LW radiation and hence shield the inner region. This makes $\text{H}_2$ a robust molecule that can exist in a UV radiation field. Previous studies point out the importance of self-shielding in preserving the amount of molecular hydrogen required for star formation. When simulating a field dwarf galaxy, \cite{Christensen+2012} noticed an increase in the amount of cold gas and a clumpier interstellar medium when \corr{incorporating non-equilibrium $\text{H}_{2}$ chemistry and self-shielding}. \cite{Safranek-Shrader+2017} also found that self-shielding is crucial in the development of $\text{H}_{2}$ in a galaxy disc's mid-plane. Therefore, the $\text{H}_{2}$ self-shielding mechanism is key to preserving the amount of $\text{H}_{2}$ and regulating star formation. 

Through modeling a semi-infinite, static slab of gas irradiated on one surface, \cite{Draine+1996} proposed an analytical expression to model $\text{H}_{2}$ self-shielding using a shield factor that is a function of the $\text{H}_{2}$ column density. In the past, computing the column density in three dimensions was prohibitively computationally expensive \citep{Shang+2010, Wolcott-Green+2011}. Therefore, multiple methods have been devised to approximate the effect and alleviate its demanding computational cost. One method is to define a local characteristic length scale across which the $\text{H}_{2}$ number density is constant. This characteristic length can be either the Jeans length \citep{Shang+2010, Johnson+2011}, the Sobolev length \citep{Sobolev+1957,Yoshida+2006}, or a Sobolev-like density-gradient length \citep{Gnedin+2009, Gnedin+2011}. Rather than a single characteristic length, \cite{Hartwig+2015} \corr{employed the TreeCol algorithm \citep{Clark+2012}} and approximated the 3D column densities by creating spherical maps of the column density around each Voronoi cell with 48 equal-area pixels and taking into account the Doppler effect from the relative velocities of the infalling gas particles. Another non-local method is the six-ray approximation \citep{Yoshida+2003,Yoshida+2007,Glover+2007, Glover+2007a}, where the $\text{H}_{2}$ column density is integrated along six directions along three Cartesian axes centered at each particle's position. Many cosmological simulations use one of these approximations to reduce computational cost, for example, the \citealp{Christensen+2012}'s simulation, the COLDSIM simulation \citep{Maio+2022}, and the Renaissance Simulations \citep{OShea+2015}. %more in the introduction of Gredin and Draine 2014 if needed

However, there are limitations to the applicability of approximate treatments in various test problems. In the context of photodissociation of $\text{H}_{2}$ in protogalaxies, \cite{Wolcott-Green+2011} showed that the Sobolev length, the density gradient, and the six-ray approximation methods underestimate $\text{H}_{2}$ self-shielding by an order of magnitude in low-density regions of $n < 10^{4} \text{cm}^{-3}$. In the context of direct-collapse black hole, when comparing with their non-local approximation, \cite{Hartwig+2015b} found that the Jeans approach returns a critical flux value $J_{crit}$ (the lowest flux required for a halo with virial temperature above $10^{4}$ K to collapse to a supermassive black hole seed) two times larger than the non-local approximation, leading to a discrepancy in the predicted number density of black hole seeds. \cite{Greif+2014} investigated the collapse of primordial star-forming clouds with an accurate $\text{H}_{2}$ line emission model and noticed that the Sobolev method brings about a thermal instability for the collapsing cloud and an order-of-magnitude overestimation of the escape fraction for high optical depth. \cite{Luo+2020} compared three self-shielding approximation models (Jeans, Sobolev-like, and 0.25$\times$Jeans) when examining the critical conditions for direct collapse black hole formation. They found that the 0.25$\times$Jeans model best approximates direct integration column density at a single sightline, while the Jeans approach overestimates and the Sobolev-like approach underestimates it. When evaluating the triggered Population III star formation at the limb of an \corrtwo{$\text{HII}$} region, \cite{Chiaki+2023a} found that while the density-gradient method matches well with the ray-tracing method, the Jeans length approximation results in an overestimation of the number of Population III stars at the front of the shock wave. In their large-scale galactic discs simulations, \cite{Safranek-Shrader+2017} noticed that the Sobolev method underpredicts the $\text{H}_{2}$ abundance in the disc by a factor of 5, while the six-ray and the Jeans length methods perform better. This is contrary to the previously mentioned studies on other problems, which found the Jeans length method to be ineffective. Thus, the validity of these approximation methods is highly problem-dependent, and they cannot be used as a general substitution for the full ray-tracing calculation. 

In this paper, we expand the investigation by comparing the Sobolev-like density-gradient approximation model \citep{Gnedin+2009} with a detailed ray-tracing method in the context of star formation during the Reionization period. This will help further examine and inform the community about the applicability of these approximations in modeling $\text{H}_{2}$ in different contexts. In Section~\ref{sec:method}, we describe our simulation, the implemented $\text{H}_{2}$ self-shielding models, and our methodology for analysis. Section~\ref{sec:result} analyzes the effect of these models on star formation and galaxy properties. This section will be subdivided into smaller sections that discuss the effect of the self-shielding model on individual galaxies and on the large-scale universe. Section~\ref{sec:comparison} discusses the comparison between our results and the literature's. Finally, in Section~\ref{sec:conclusion}, we discuss the implications of the findings and summarize the paper.

%\textit{Maybe a paragraph at the beginning about the observational and theoretical evidence of the relation between H2 and star formation (Kenicutt-Schmidt relation, for example)?}\acorr{I think this is a good idea. I'm happy to help with this part if you need some background.}

\section{Methods} \label{sec:method}
\subsection{Comoslogical Simulations}
\label{subsec:cosmological_simulations}
We ran and analyzed outputs from a set of cosmological radiation-hydrodynamic adaptive mesh refinement ENZO \citep{Bryan+2014} simulations. Four cosmological simulations were generated to explore the self-shielding effect in the early stage of the Reionization period, which are referred to as ``\textit{ray}'', ``\textit{apx1}'', ``\textit{apx2}'', and ``\textit{apx-c}''. The ``\textit{ray}'' simulation uses the adaptive ray-tracing method \citep{Wise+2011} for the LW photon and $\text{H}_{2}$ self-shielding. The other three simulations, ``\textit{apx1}'', ``\textit{apx2}'', and ``\textit{apx-c}'', assume an optically-thin LW radiation field and use a local approximation of the $\text{H}_{2}$ column density when computing the $\text{H}_{2}$ self-shielding factor. For convenience, we use the term ``\textit{apx}'' to refer to those three simulations together. The detailed implementations and differences between those simulations will be discussed in Section~\ref{sec:h2_selfshielding_models}. All four simulations include non-equilibrium, radiatively driven chemistry with a network of 9 species: \corr{H, $\text{H}^{+}$, He, $\text{He}^{+}$, $\text{He}^{++}$, $e^{-}$, $\text{H}_{2}$, $\text{H}_{2}^{+}$, and $\text{H}^{-}$}. The network includes $\text{H}_{2}$ formation through the three-body channel ($\text{H} + \text{H} + \text{H} \rightarrow \text{H}_{2} + \text{H}$), the collisional $\text{H}^{-}$ channel ($\text{H} + e^{-} \rightarrow \text{H}^{-} + \gamma$ and $\text{H}^{-} + \text{H} \rightarrow \text{H}_{2} + e^{-}$) and the $\text{H}^{+}$/$\text{H}_{2}^{+}$ channel ($\text{H} + \text{H}^{+} \rightarrow \text{H}_{2}^{+} + \gamma$ and $\text{H}_{2}^{+} + \text{H} \rightarrow \text{H}_{2} + \text{H}^{+}$). The excitation of $\text{H}_{2}$ and the dissociation of $\text{H}_{2}$ and $\text{H}_{2}^{+}$ molecules through radiative processes are also implemented in the network. \corr{We also track metallicity and enable metal cooling that uses heating/cooling values from \cite{Smith+2008}, which was created with \textsc{CLOUDY} \citep{Ferland+1998}}. The four simulations are initialized with an identical list of parameters and initial conditions, except for the parameters pertaining to the LW radiation field.

Depending on the number of active star particles and the refinement levels in our simulation box at a given timestep, it may take 2 to 10 times longer to advance to the next timestep when using ray-tracing. For instance, when running with 400 cores, 3GB of memory each core, on the Illinois Campus Cluster Intel Cascade Lake node to evolve the simulations from $z = 12.02$ to $z = 12$, it took about 12 hours to run the ``\textit{ray}'' simulation while it only took 2.5 hours for the ``\textit{apx}'' simulations. Due to this higher computational cost of the ``\textit{ray}'' simulation, we only evolved it from $z = 100$ to $z = 12$ while we evolved the other three simulations down to $z \approx 10.89$. Among all four simulations, the first star particle is formed earliest in the ``\textit{apx-c}'' simulation, which happens 206 million years after the Big Bang. To capture the dynamics of star formation and feedback,  we output snapshots from each simulation every 1 million years, starting from 20.5 million years after the Big Bang (one snapshot before the first snapshot with stars). We use the analysis toolkit \texttt{yt} \citep{Turk+2011} to load the simulation and visualize volumetric data. 
%10 times longer is at DD0293 to DD0294 for example

All simulations assume a flat $\Lambda$CDM cosmology and are run with the cosmological parameters obtained from \cite{PlanckCollaboration+2016}: $\Omega_{M} = 0.3065$, $\Omega_{\Lambda} = 0.6935$, $\Omega_{b} = 0.0483$, $h = 0.679$, $\sigma_{8} = 0.8154$, and $n = 0.9681$. The total moving volume is 13 $\text{(Mpc/h)}^{3}$ and the root grid dimension is $256^{3}$. Each simulation contains $\approx 10^{7.8}$ dark matter particles. Inside the total volume, we create a smaller zoom-in region using three additional levels of nested refinement to create a grid with size $\approx 220 \times 220 \times 240 \; \text{kpc}^{3}$ (at $z = 10.89$) with an effective grid size resolution of $2048^{3}$. In the zoom-in region, the most refined dark matter particle mass is $2.7\times 10^{4}$ $M_\odot$. If a grid inside this zoom-in region has the baryon mass above $\approx 10^{7} M_\odot$ or the dark matter particle mass above $\approx 5.5\times10^{7} M_\odot$, that grid can further be refined adaptively up to $2^{-13}$ times the root grid dimensions. This corresponds to a maximum spatial resolution of 0.77 pc at $z = 10.89$ (or $0.7$ pc at $z = 12$), which allows the tracking of large molecular clouds through initial collapse. We use radiating Population II (PopII) star cluster particles \citep{Wise+2009} and Population III (PopIII) star particles \citep{Abel+2007} for our star formation and radiative feedback prescription. 

To identify the halos in the simulations, we employ \textsc{Haskap Pie} \citep{Barrow+2025}, an all-in-one algorithm that both finds halos and builds merger trees using overdensity-finding, energy-solving, cluster-finding, and particle tracking. \corr{Thanks to the energy-based halo solutions, a halo found by \textsc{Haskap Pie} does not need to be spherical and only comprises dark matter particles that are bound to each other.} We exclude all halos that are outside the refined region of the simulation box to ensure that the halos contain only refined particles. We also enforce that a halo found by \textsc{Haskap Pie} must have at least 5 dark matter particles. \corr{However, in practice, the smallest halo has 7 particles, over 95\% of all our halos across all snapshots have more than 50 particles ($M_\text{halo} > 1.4\times10^{6} M_\odot$), and over 99\% of our halos have more than 30 particles ($M_\text{halo} > 8\times10^{5} M_\odot$). Given their limited quantity, very small halos do not affect our overall conclusions.}

\subsection{Numerical \texorpdfstring{$\text{H}_{2}$}{H2} Self-shielding Models}
\label{sec:h2_selfshielding_models}

\begin{deluxetable*}{@{\extracolsep{3em}}l|cccc}
\tablecaption{ENZO LW Radiative Transfer Parameters \label{tab:enzo_params}}
\tablehead{
\colhead{ENZO parameter} & \colhead{\textit{ray}} & \colhead{\textit{apx1}} & \colhead{\textit{apx2}} & \colhead{\textit{apx-c}}
}
\startdata
\texttt{RadiativeTransferOpticallyThinH2} & 0 & 1 & 1 & 1 \\
\texttt{RadiativeTransferUseH2Shielding} & 1 & - & - & - \\
\texttt{RadiativeTransferH2ShieldType} & 1 & - & - & - \\
\texttt{RadiationShield} & 0 & 2 & 2 & 2\tablenotemark{*} \\
\enddata
\tablenotetext{*}{The formula for the approximated self-shielding column density and self-shielding factor was corrected in the file \texttt{solve\_rate\_cool.F} of ENZO's stable version.}
\end{deluxetable*}

An $\text{H}_{2}$ self-shielding process happens when a molecular cloud has a sufficiently high $\text{H}_{2}$ column density that its outer layer absorbs incoming LW UV photons from nearby stars, effectively shielding the molecules within it against photodissociation. Via modeling a semi-infinite, static slab of gas that is irradiated on one surface, \cite{Draine+1996} expressed a self-shielding-accounted photodissociation rate as
\begin{equation}
    k_{\text{diss}}(N_{\text{H}_{2}},T) = f_{sh}(N_{\text{H}_{2}},T)\cdot k_{\text{diss}}(N_{\text{H}_{2}}=0,T),
\end{equation}
where $f_{sh}(N_{\text{H}_{2}},T)$ is a self-shielding factor or shielding function, and $k_{\text{diss}}(N_{\text{H}_{2}}=0,T)$ is the dissociation rate in an optically-thin regime. It is important to note that the convention of the $f_{sh}$ value is slightly counterintuitive: the $f_{sh}$ value of 0 means the gas cell is fully shielded from LW radiation, and the $f_{sh}$ value of 1 means that there is no self-shielding. The self-shielding factor can be approximated using the fitting function from \citet{Wolcott-Green+2011} that incorporates a temperature dependence due to Doppler thermal broadening, 
\begin{equation}
\begin{aligned}
    f_{sh} = \frac{0.965}{(1+x/b_{5})^{1.1}}& +  \frac{0.035}{(1+x)^{0.5}} \\ 
    &\times \exp{[-8.5\times 10^{4}(1+x)^{0.5}]},
\label{eq:self_shielding_factor_tempeq}
\end{aligned}
\end{equation}
where $x = N_{\text{H}_{2}}/(5 \times 10^{14}\;\mathrm{cm}^{-2})$, $b_{5} = b/(10^{5} \text{cm}\,\text{s}^{-1})$, $b = \sqrt{2k_{B}T/m_{\text{H}_{2}}}$, \corr{and $T$ is the gas cell temperature}. \corr{This fitting function is based closely yet improves upon the \cite{Draine+1996}'s fitting function by adjusting the exponent of the temperature term $(1+x/b_{5})$ from 2 to 1.1 to provide a better fit in the high temperature regime.} This fitting is most accurate for moderate gas densities ($n \approx 10^{3} \mathrm{cm}^{-3}$) and temperatures ($T \approx 10^{3} \text{K}$) \citep{Wolcott-Green+2011}. Recently, \cite{Wolcott-Green+2019} introduced a newer fit that can be applicable up to $n \approx 10^{7} \text{cm}^{-3}$ and $T \approx 8\times10^{3} \text{K}$. However, this fitting formula has not been implemented in ENZO's LW radiative transfer pipeline yet by the time this paper is written. Comparisons using this formula will be examined in a future project. The ENZO's LW radiative transfer parameters of each simulation are listed in Table~\ref{tab:enzo_params}.

In this paper, we compare results from two methods: a direct integration from adaptive ray-tracing (by setting \texttt{RadiativeTransferOpticallyThinH2} = 0) and a Sobolev-like length approximation \citep{Gnedin+2009, Wolcott-Green+2011}(by setting \texttt{RadiativeTransferOpticallyThinH2} = 1 and \texttt{RadiationShield} = 2). For the ray-tracing method, we use the adaptive ray-tracing module \textsc{MORAY} \citep{Wise+2011} that is based on the HEALPix (Hierarchical Equal Area isoLatitude Pixelation, \citealt{Gorski+2005}) scheme. For each simulation cell, the $\text{H}_{2}$ dissociation rate is calculated by,
\begin{equation}
    k_\mathrm{diss} = \sum _{\rm rays} 
\frac{ P_{\rm LW} \sigma _{\rm H_2} \Omega _{\rm ray} r^2 \mathrm{d} r }
{ A_{\rm cell} V_{\rm cell} \mathrm{d} t_{\rm P}},
\end{equation}
where \corr{$P_{\rm LW}$ is the LW photon number flux at the entry of the cell into the ray, $\Omega _{\rm ray}$ is the solid angle of a {\sc HEALPix} cell, $r$ is the distance between a source and the cell, $A_{\rm cell}$ is the cell's area, $V_{\rm cell}$ is the cell's volume, $\mathrm{d} t_{\rm P}$ is the photon integration timestep, and $\sigma _{\rm H_2} = 3.71\times10^{-18}~{\rm cm}^{2}$ is the effective cross-section area of $\text{H}_{2}$ against LW photons that accounts for the probability that a given absorption is followed by dissociation \citep{Abel+1997}}. Through each cell, the LW photon is attenuated by
\begin{equation}
    \mathrm{d} P_{\rm LW} = P_{\rm LW}
\left(1 - \frac{f_{\mathrm{sh}}(N_{\text{H}_{2}}+\mathrm{d}N_{\text{H}_{2}})}{f_{\mathrm{sh}}(N_{\text{H}_{2}})}\right),
\end{equation}
\corr{where $\mathrm{d}N_{\text{H}_{2}}$ is the $\text{H}_{2}$ column density in the current cell}. The column density is integrated along a HEALPix ray along the line element $ds$,
\begin{equation}
    N_{\text{H}_{2}} = \int n_{\text{H}_{2}}\text{d}s.
\label{eq:NH2_integrate}
\end{equation}

The Sobolev-like density-gradient treatment \citep{Gnedin+2009, Gnedin+2011} is a local approximation in which the number density $n_{\text{H}_{2}}$ is assumed to be constant within a characteristic length scale $L_{\text{char}}$. In this approximation, the characteristic length scale represents the distance over which the cell's gas density, $\rho$, diminishes. The gas beyond this distance is sparse enough that its influence on shielding is negligible, and therefore $\text{H}_{2}$ is dissociated in an optically thin fashion. The column density is then computed by
\begin{equation}
    N_{\text{H}_{2}} \approx n_{\text{H}_{2}}L_{\text{char}} =  n_{\text{H}_{2}}\frac{\rho}{2|\nabla\rho|}.
\label{eq:NH2_Sobolev-like_approx}
\end{equation}
Because we assume that the gas beyond the characteristic length is optically thin for LW radiation, we only reduce the photon emission rates from the source to the gas cell by the inverse square law without any additional attenuation from the medium in between. The $\text{H}_{2}$ photodissociation rate in the approximation method is thus computed by
\begin{equation}
k_\mathrm{diss} = f_{sh}(N_{H_2})
\frac{Q_{\mathrm{LW}} \sigma _{\rm H_2}}{4\pi r^2}.
\label{eq:kdiss_apx}
\end{equation}
The LW photon emission rate $Q_{\mathrm{LW}}$ in photons per second is calculated by
\begin{equation}
  Q_\mathrm{LW}(m_{\bigstar})=\begin{cases}
    1.12\times10^{46}m_{\bigstar}\times\chi_{\text{LW}}(1-\chi_{\text{HeI}}-\chi_{\text{HeII}}),\\ \qquad\qquad\qquad\qquad \text{if \;PopII < 20 Myr}, \\
    10^{44.03 + 4.59m_{\bigstar}  - 0.77m_{\bigstar}^{2}},\qquad\quad \text{if\; PopIII}
  \end{cases}
\label{eq:QLW}
\end{equation}
where $m_{\bigstar}$ is the mass of the star particle in solar mass. $\chi_{\text{LW}}$, $\chi_{\text{HeI}}$, and $\chi_{\text{HeII}}$ are the ratios between the energy output of LW photons, $\text{He}$ ionizing photons, and $\text{He}^{+}$ ionizing photons with respect to to H ionizing photons. $\chi_{\text{LW}}$, $\chi_{\text{HeI}}$, and $\chi_{\text{HeII}}$ are set to 1.288, 0.2951, and $2.818\times10^{-4}$ \citep{Schaerer+2003}. Furthermore, because LW radiation is mainly emitted by young massive stars that are short-lived, only PopII star cluster particles younger than 20 Myr are modeled to emit LW photons in our simulations. For PopIII stars, the lifetime of a PopIII star particle is between 1.8 to 62 Myr, depending on its mass. The LW photon emission rate of PopIII stars is taken from \cite{Schaerer+2002}.

During the course of our investigation, we identified and implemented a correction to the ENZO's \texttt{solve\_rate\_cool.F} file, which governs the calculation of the self-shielding factor for the approximation method. Specifically, in the earlier version of the code, the thermal broadening parameter $b$ in Equation~\ref{eq:self_shielding_factor_tempeq} is calculated with $m_{H}$ instead of $m_{\text{H}_{2}}$ and the calculation of the approximated $\text{H}_{2}$ column density in Equation~\ref{eq:NH2_Sobolev-like_approx} missed a factor of 2 in the denominator. The corrected codes are now approved to merge with the ENZO code\footnote{\url{https://github.com/enzo-project/enzo-dev/pull/246}}.

In summary, the simulation ``\textit{ray}'' uses ray-tracing to solve radiative transfer equations of LW radiation; ``\textit{apx1}'' and ``\textit{apx2}'' assume an optically thin LW radiation field and the Sobolev-like approximation to estimate $\text{H}_{2}$ self-shielding; and ``\textit{apx-c}'' also includes an optically-thin LW assumption as ``\textit{apx1}'' and ``\textit{apx2}'', but has corrected self-shielding formulae in the numerical implementation. It is important to note that ``\textit{apx1}'' and ``\textit{apx2}'' simulations are run with the exact same set of parameters but their outcomes slightly differ from each other due to the random sampling of PopIII stellar mass during its formation \citep{Bryan+2014}. Our PopIII routine selects masses randomly sampled from an initial mass function and then assigns a supernova feedback strength based on the selected mass. Therefore, each realization of a simulation will be subject to a different amount of PopIII supernova feedback, even if initial conditions are identical. These differences help us explore the range of outcomes that the approximation model can return and make comparisons with the \textit{ray} simulation more robust. Despite these differences, the global trend in the \textit{apx} simulations remains consistent. 

\subsection{Cross-matching Halos}
\label{subsect:cross-matching_halos}
To examine how each individual halo is affected by the implementation scheme of $\text{H}_{2}$ self-shielding, we perform halo cross-matching between simulations. The matching requirements are
\begin{align}
    & \frac{2}{3} < \frac{M_{\mathrm{halo}_{\mathrm{ray}}}}{M_{\mathrm{halo}_{\mathrm{apx}}}} < \frac{3}{2}, \nonumber \\
    \frac{d_{\text{center}}}{R_{\mathrm{halo}_{\mathrm{ray}}}} & < 0.5 \; , \; \frac{d_{\text{center}}}{R_{\mathrm{halo}_{\mathrm{apx}}}} < 0.5,
\label{eq:crossmatching_requirement}
\end{align}
where $R_{\mathrm{halo}} = R_{200c}$ is the radius containing the dark matter density 200 times the universe's critical density, $M_{\mathrm{halo}} = M_{200c}$ is the dark matter mass enclosed within $R_{200c}$, and $d_{\text{center}}$ is the distance between the halo centers. Because we are mainly interested in main progenitor halos that host stars, and these halos are typically the most massive in our volume ($M_\text{halo} \geq 10^{9}M_\odot$ at $z = 12$), these simple matching requirements are sufficient to find counterpart halos among the simulations without further sophisticated matching methods. We cross-match all the halos at the last timestep of the ``\textit{ray}'' simulation ($z = 12$) and then trace back their merger trees.

\subsection{Defining ISM, CGM, and IGM}

Because the approximation model depends on gas density and temperature (Equations~\ref{eq:self_shielding_factor_tempeq} and \ref{eq:NH2_Sobolev-like_approx}), its effect may differ between different parts of the simulated universe. Thus, we partition our volume into three regions: the interstellar medium (ISM), the circumgalactic medium (CGM), and the intergalactic medium (IGM) region. In our analysis, the boundary of these regions is defined as follows. The ISM region (or the boundary of a galaxy) starts from the galaxy's center of mass to $R_{\mathrm{bary}, 2000}$, which is the radius enclosing a baryonic (gas and stars) mass density 2000 times the universe's critical mass density at a given redshift. Even though there are multiple other definition of the boundary of galaxy in the literature, such as using a fixed fraction of the virial radius, a fixed physical radius, or a radius that is scaled with mass or surface brightness (see \citealp{Stevens+2014} for a summary of these techniques), our over-density approach helps define the galaxy more robustly and more physically because it is redshift and density dependent. It also helps better define the galaxy during a close galaxy interaction when a half-mass radius or half-light radius approach may fail as another galaxy is in the target galaxy's virial radius. Depending on the redshift and the galaxy's compactness, $R_{\mathrm{bary}, 2000}$ typically ranges from 0.15 to 0.25 times the halo's $R_{200c}$, which falls in the same range as other galaxy aperture definitions in cosmological simulations \citep{Stevens+2014}. For the rest of the paper, we use the term "galaxy" to refer to a region within $R_{\mathrm{bary}, 2000}$ of our halo. The CGM starts from $R_{\mathrm{bary}, 2000}$ to $5R_{\mathrm{bary}, 2000}$, which is approximate to the halo's $R_{200c}$. The IGM is the region outside of the $R_{200c}$ of all halos.

Also, it is important to note that the baryonic center of mass does not always coincide with the dark matter center of mass, especially during halo interactions. Thus, for our galaxy analysis, we center our galaxy on the baryonic (gas plus star) center of mass instead of the dark matter center of mass. The baryonic center is determined by an iterative method where we choose an initial center, then we iteratively expand out and redefine the center until a certain baryonic overdensity is reached. This initial center is chosen to be the center of mass of the inner-radius stars (for the last timestep of the simulation) or the baryonic center found in a previous snapshot (for the rest of the snapshots). This shift in the galactic center helps locate the ISM and CGM better and allows us to examine the radial profile of a galaxy's properties more accurately. 
%Because stars are principally located deep in the halo's potential well, the shift is small compared to the halo's $R_{200c}$ and does not move the galaxy outside the halo. (REMOVED BECAUSE THE SHIFT CAN BE LARGE DURING GALAXY INTERACTION, AS IN THE CASE WITH GIZMO AND RAMSES IN THE AGORA DATASET). 

\subsection{Star Assignment}
\label{subsect:star_assignment}

We uniquely assign each star particle in the simulation to at most one dark matter halo in a two-step fashion. For this procedure, a halo's boundary is defined as $R_{200c}$. For the first step, a star particle is assigned to a halo where it is first created. If a star particle is created inside the intersection of multiple halos, we will calculate the orbital energy of that star particle with respect to the dark matter particles within $R_{200c}$ of each halo to determine which halo the particle belongs to. The assigned halo is the one with a lower negative relative orbital energy. If a star has positive total orbital energy with respect to all halos it is in (i.e., it is bound to no halos), we keep computing its orbital energy for the rest of the timesteps and assign it to a halo if the energy is negative. In this first step, we assume that a star particle never leaves its assigned halo unless that halo merges with another one, in which case the star particle will become a member of the descendant halo of the merger. This assumption allows a quick assignment of stars to halos without the need to calculate the orbital energy of each star with respect to multiple halos, which is computationally expensive. The assignment's first step starts from the first snapshot to the last snapshot of the simulation. 

Even though stars do not generally leave a halo's potential well for isolated halos, the behavior becomes more complex during halo-halo interaction as stripping can occur and stars can be lost from one halo to another \citep{Kannan+2015}. Therefore, the second step of our assignment process is to address this shortcoming of the first step's assumption and to re-evaluate the star assignment result. In this step, we validate the output from the first step and check whether a star remains in its assigned halo's $R_{200c}$ throughout its lifetime. If a star escapes the radius $R_{200c}$ of its originally assigned halo, we assign that star to a new halo under two conditions: (1) the star's position must be within the new halo's $R_{200c}$ and (2) the star's total orbital energy with respect to the new halo must be negative. Similar to the first step, if a star is located inside multiple possible new halos, we assign that star to the halo with the lowest negative orbital energy. If either condition (1) or (2) fails at a certain timestep, this suggests that the star is stripped out of the potential well of all halos, and thus, we remove that star from the halo assignment of that time step. Also, after a star gets out of its originally assigned halo, even if it comes back to that halo later in the following timesteps, it still needs to satisfy the above two conditions to be reassigned back to that halo. This helps us avoid incorrect assignments in situations where stars get unbound from a halo and then just happen to fly through that halo later. For context, we find that stellar stripping during halo-halo interaction can occur up to 20\% of all assigned star particles in the box, emphasizing the necessity of this re-assignment step. 

Once each star particle is uniquely associated with at most one halo, we calculate a halo's stellar mass and star formation rate (SFR) exclusively based on its assigned star particles. This gravitational unbinding of star particles guarantees that each halo's stellar mass profile does not overlap and is independent of the others. 
 
\section{Results}
\label{sec:result}

\subsection{Comparison of the \texorpdfstring{$\text{H}_{2}$}{H2} Photodissociation Rate}

\begin{figure*}
	\centering
	\includegraphics[width=1\textwidth]{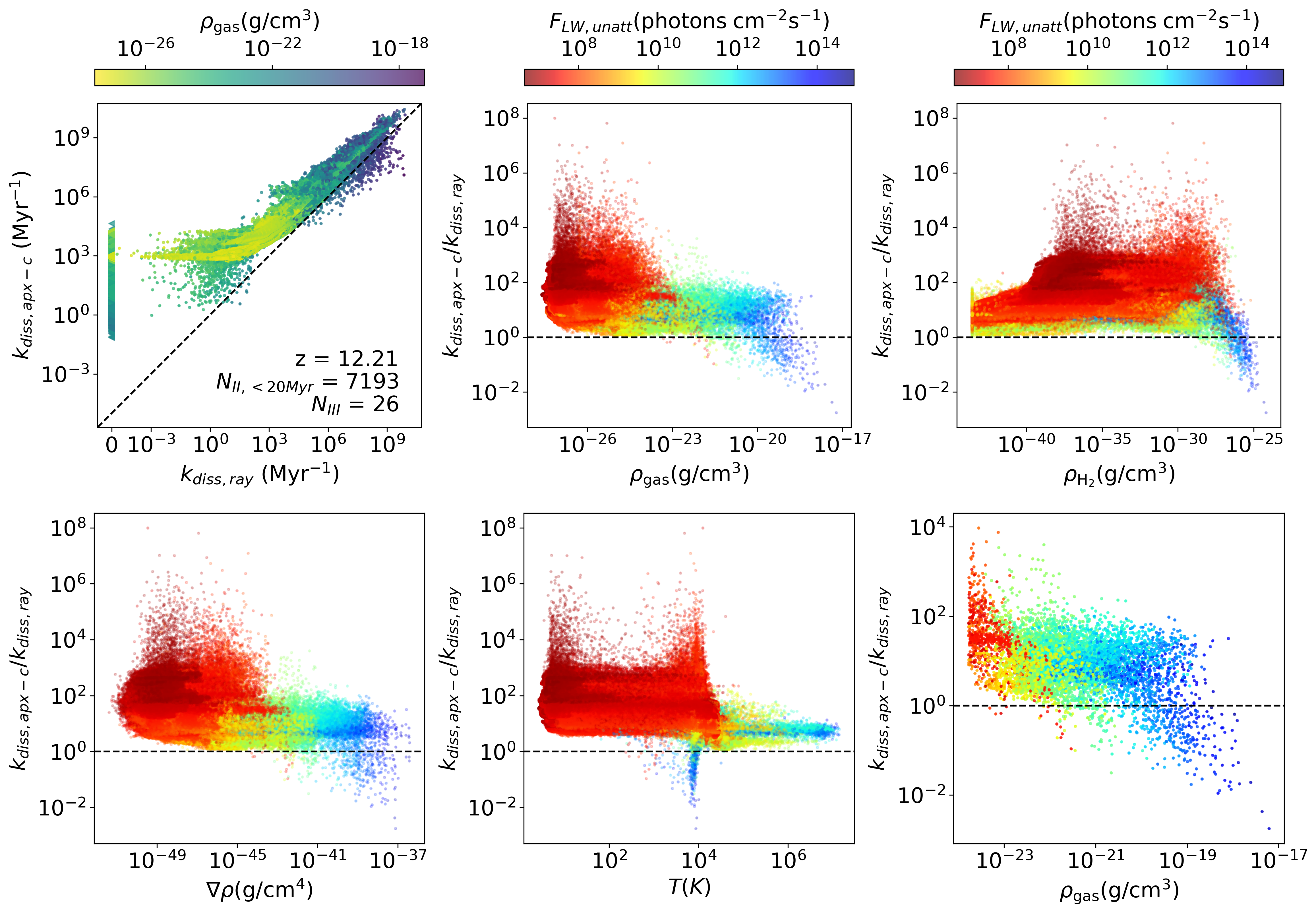}
	\caption{Comparison between the $\text{H}_{2}$ photodissociation rate obtained from the ``\textit{ray}'' simulation ($k_\mathrm{diss, ray}$) and the rate manually calculated by using the Sobolev-like approximation ($k_\mathrm{diss,apx-c}$). The data points represent 
    all gas cells in a region surrounding $5R_{200c}$ of all halos with stars. (Top left) The relationship between $k_\mathrm{diss, ray}$ and $k_\mathrm{diss,apx-c}$, colored by the cell's gas density. The vertical streak on the left side of the subplot represents gas cells whose $k_\mathrm{diss} = 0$ is in the ``\textit{ray}'' simulation but is non-zero using the approximated calculation. (Top middle, top right, bottom left, bottom \corr{middle}) The ratio between the approximated and the ray-traced photodissociation rate plotted as a function of gas density, $\text{H}_{2}$ density, gas density gradient magnitude, and gas temperature, colored by the total unattenuated LW flux received by a cell. \corr{(Bottom right) Similar to the top middle plot, but we focus on the gas cells with density and $\text{H}_{2}$ whose $\text{H}_{2}$ cooling is important}. The Sobolev-like approximation tends to \corr{enhance} the amount of $\text{H}_{2}$ self-shielding in the high gas density regime (thus lower $k_\mathrm{diss}$) and \corr{weaken} it in the low gas density regime (thus higher $k_\mathrm{diss}$).}
	\label{fig:kdiss_apxc-ray_comparison}
\end{figure*} 

\begin{figure}
	\centering
	\includegraphics[width=0.8\columnwidth]{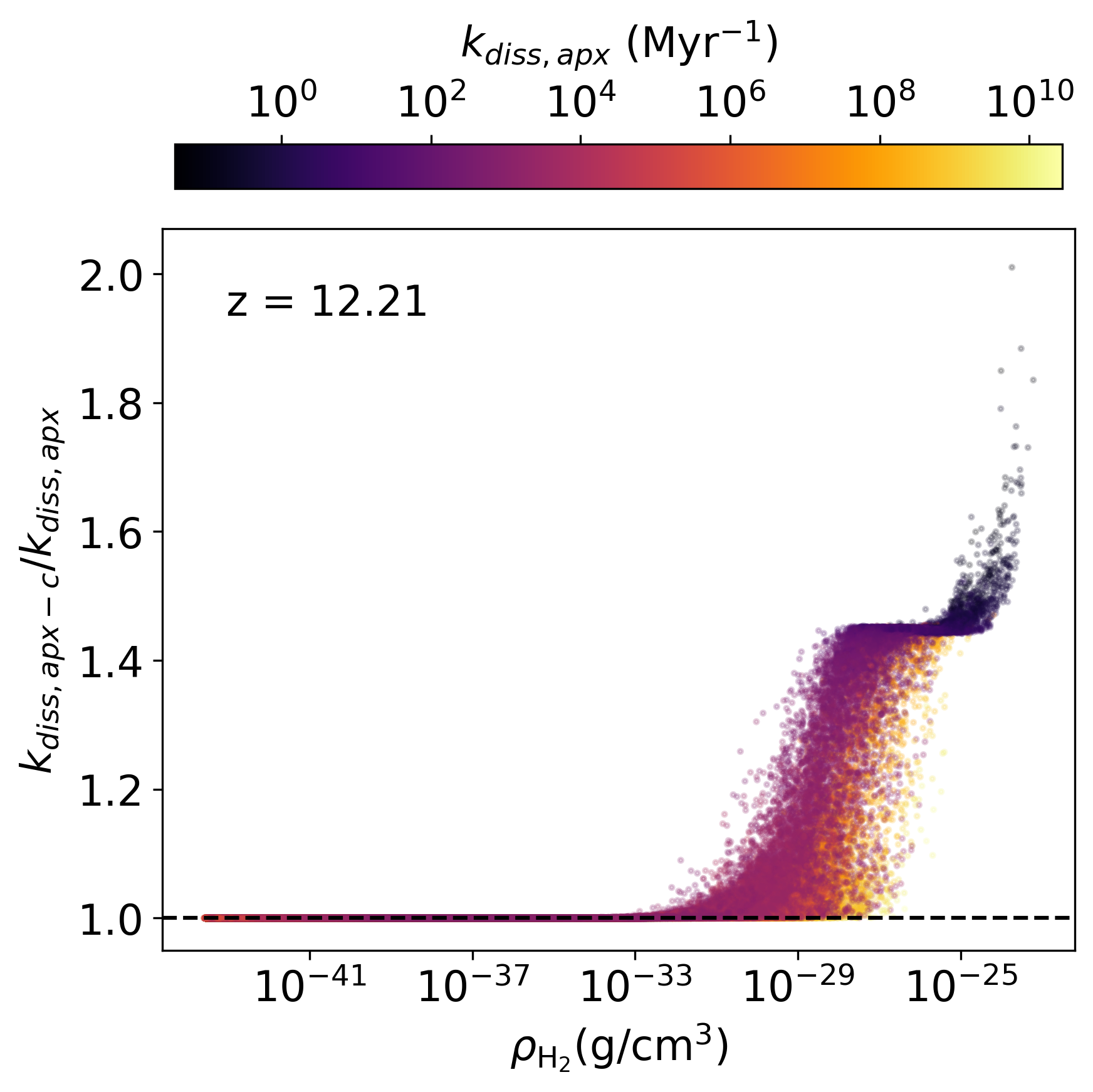}
	\caption{The ratio between the old ENZO implementation of the Sobolev-like approximation and the corrected implementation as a function of $\text{H}_{2}$ density. The rates are manually calculated using the same set of gas cells in Fig.~\ref{fig:kdiss_apxc-ray_comparison}. The color represents the value of the old approximated $k_{\mathrm{diss}}$. We observe that the correction does not significantly change the photodissociation rate calculation.}
	\label{fig:kdiss_apxc-apx_comparison}
\end{figure} 

Fig.~\ref{fig:kdiss_apxc-ray_comparison} compares the $\text{H}_{2}$ photodissociation rate $k_\mathrm{diss}$ between the ray-tracing model and the Sobolev-like approximation model with the corrected ENZO implementation. For every gas cell in \corr{a $\approx 5\times10^{-4}\;\text{Mpc}^{3}$ region surrounding all halos with stars at $z = 12.21$}, we obtained the photodissociation rate of each cell in the ``\textit{ray}'' simulation (which was already calculated by ENZO during the simulation run and can be loaded by \texttt{yt}) and manually recalculated that rate by using the Sobolev-like approximation method (Equations~\ref{eq:NH2_Sobolev-like_approx} and \ref{eq:kdiss_apx}). To compare the instantaneous effect of the approximation on the same set of gas cells, we manually computed the approximated rate on the ``\textit{ray}'' simulation's cells instead of using the cells in any of the \textit{apx} simulations. Also, it is more complex to cross-match every single gas cell between the four simulations because the cells can be refined differently in each simulation. We choose the timestep at $z = 12.21$ to show the comparison because this timestep has gas cells covering a large range of gas density (from $10^{-27}\;\mathrm{g}/\mathrm{cm}^{3}$ up to $10^{-17}\;\mathrm{g}/\mathrm{cm}^{3}$), but other timesteps also show a similar behavior. The top left plot of Fig.~\ref{fig:kdiss_apxc-ray_comparison} demonstrates a clear discrepancy in the $k_{\mathrm{diss}}$ between the two radiative models. The approximation matches better with the more accurate ray-traced solution for gas cells with a high photodissociation rate while struggling with cells whose rate is low. As shown in the top left subplot, $k_\text{diss}$ is higher in denser cells. This is because star particles only form in dense simulation cells. Therefore, higher-density gas cells are more likely to be located near active star particles and thus more exposed to LW radiation, leading to a higher photodissociation rate. When we examined gas density, we found that the approximation model tends to \corr{intensify} the amount of shielding for cells with a gas density larger than $10^{-20} \text{g}/\text{cm}^3$ and can under-predict the photodissociation rate by up to two order of magnitude (2 dex), as shown in the top left and top middle subplots of Fig.\ref{fig:kdiss_apxc-ray_comparison}. For a fixed density gradient, a higher gas density means a higher self-shielding characteristic length (Equation~\ref{eq:NH2_Sobolev-like_approx}). If a star particle is located within a cell but the self-shielding characteristic length is larger than the cell's size, the $\text{H}_{2}$ column density will be exaggerated by the approximation model, leading to more shielding. In summary, the higher the gas density above $10^{-21} \text{g}/\text{cm}^3$, the more likely the approximated $k_\mathrm{diss}$ is smaller compared to the ray-traced $k_\mathrm{diss}$.

According to Equations~\ref{eq:self_shielding_factor_tempeq} and \ref{eq:NH2_Sobolev-like_approx}, the approximated column density for self-shielding also depends on the $\text{H}_{2}$ density of the cell, the gas density gradient magnitude, and the gas temperature. We explore the dependency of the approximation model's accuracy on all these variables. For a given gas density and gas density gradient, a cell with a higher amount of $\text{H}_{2}$ has a larger approximated self-shielding column density (Equation~\ref{eq:NH2_Sobolev-like_approx}) and hence a smaller self-shielding factor (Equation~\ref{eq:self_shielding_factor_tempeq}). As a result, cells with high $\text{H}_2$ density ($\rho_{\text{H}_2} > 10^{-27}\,\text{g}/\text{cm}^{3}$) have a lower rate of photodissociation when the approximation is used (top right subplot of Fig.~\ref{fig:kdiss_apxc-ray_comparison}). This can result in a runaway effect. When a cell passes a certain $\text{H}_{2}$ density, it becomes overshielded when using the approximation treatment. This overshielding in return makes it more difficult to dissociate $\text{H}_{2}$ when more $\text{H}_{2}$ builds up, which further drives the $\text{H}_{2}$ density higher. The increase in $\text{H}_{2}$ then makes the cell even more overshielded, potentially leading to numerically-enhanced self-shielded $\text{H}_{2}$ clumps. The relationship between $k_\mathrm{kdiss,apx-c}/k_\mathrm{kdiss,ray}$ and $\rho_{\text{H}_{2}}$ (top right subplot) is relatively tight at $\rho_{\text{H}_{2}} > 10^{-28} \text{g}/\text{cm}^{3}$, showing that $\text{H}_{2}$ can also be used to estimate the accuracy of the approximation at higher $\text{H}_{2}$ density.

The $k_\mathrm{diss}$ ratio also shows a slight dependency on the gas density gradient magnitude and gas temperature. The ratio gets smaller at larger density gradients (bottom left subplot of Fig.~\ref{fig:kdiss_apxc-ray_comparison}) and at higher temperatures (bottom right subplot of Fig.~\ref{fig:kdiss_apxc-ray_comparison}). The slight dependency of $k_\mathrm{diss}$ on gas density gradient and gas temperature is likely due to the dependence of these variables on gas density. Gas cells with higher density often have a steeper gas density gradient and are often subjected to heating by stellar feedback because stars mostly form in dense gas cells. However, we do not observe a distinct relationship between $k_\mathrm{kdiss,apx-c}/k_\mathrm{kdiss,ray}$ and $\rho_{\text{H}_{2}}$, $\nabla\rho$, or $T$ at a lower tail of the range of these variables. To sum up, among the four variables that we investigate, gas density seems to be the most important factor in determining the accuracy of the approximation because the trend is clearest for gas density: the higher the gas density, the lower the approximated photodissociation rate with respect to the ray-traced rate. 

For diffuse gas, $\text{H}_{2}$ self-shielding is \corr{reduced} when the approximation model is used. When the density is below $10^{-22} \text{g}/\text{cm}^3$, the photodissociation rate computed by the approximation is always larger than the ray-traced value. Indeed, the median value and the average value of $k_\mathrm{diss,apx-c}/k_\mathrm{diss,ray}$ is 40 and 270, respectively. For highly diffuse gas ($\rho_\mathrm{gas} \approx10^{-26} g/cm^3$), the approximated $k_\mathrm{diss}$ may be \corr{higher} by up to 8 dex. However, it is important to note that because low-density gas tends to have low $k_\mathrm{diss}$ and high-density gas tends to have high $k_\mathrm{diss}$, the degree of overestimation or underestimation does not always reflect the actual amount of $\text{H}_{2}$ being destroyed. For instance, a 6-dex difference between $k_\mathrm{diss} = 10^3\;\mathrm{Myr}^{-1}$ and $k_\mathrm{diss} = 10^{-3}\;\mathrm{Myr}^{-1}$ in a gas cell with very little $\text{H}_{2}$ does not change the result much because there is not a lot of $\text{H}_{2}$ to be photodissociated to begin with. 

\corr{The bottom right plot of Fig.~\ref{fig:kdiss_apxc-ray_comparison} limits the range of the top middle plot by showing the $k_\mathrm{diss}$ comparison of gas cells with $n_\mathrm{gas} > 1\;\text{cm}^{-3}$ ($\rho_\mathrm{gas} > 1.67\times10^{-24} g/cm^3$) and $n_\mathrm{\text{H}_{2}} > 5\times10^{-7}\;\text{cm}^{-3}$ ($\rho_\mathrm{\text{H}_{2}} > 1.67\times10^{-30} g/cm^3$). In these ranges, $\text{H}_{2}$ is a more significant coolant \citep{Glover+2012} and thus the accuracy of the self-shielding treatment in this regime is more important. Even in this more restricted range, $k_\mathrm{diss}$ can still be off by two to four orders of magnitude between the two self-shielding treatments.}

%Furthermore, for diffuse gas cells whose density is lower than $10^{-23} \text{g}/\text{cm}^{3}$, the approximation's accuracy seems to be more influenced by gas density than $\text{H}_{2}$ density, gas density gradient, or gas temperature. Cells with a large $k_\mathrm{diss} $ overestimation from the approximation ($k_\mathrm{kdiss,apx-c}/k_\mathrm{kdiss,ray} > 10^{4}$) are more distributed at low gas density than low $\text{H}_{2}$ density. Given that the self-shielding characteristic length is a function of the gas density and not the $\text{H}_{2}$ density, at low gas density, the self-shielding characteristic length becomes smaller. Thus, the effective self-shielding column density is reduced, making the gas less shielded.

Even at a given $\rho_\mathrm{gas}$ or $\rho_{\mathrm{H}_{2}}$, there is a good scatter in the difference between $k_\mathrm{diss,apx-c}$ and $k_\mathrm{diss, ray}$. This can be explained by how much of the unattenuated LW radiation flux a cell receives ($F_\text{LW, unatt}$), which is calculated by using the inverse square law on the LW photon emission rate (Equation~\ref{eq:QLW}) and is shown by the color of the scatter points. 
%can be roughly quantified by the distance from that cell to the nearest active PopII star particle (age younger than 20 Myr). We decide to use the distance to the nearest PopII star particle instead of any star particle because a PopII star particle ($M_\text{PopIII} \approx 10^{3}-10^7M_\odot$ in our simulation) is much more massive than a PopIII star particle ($1M_\odot \leq M_\text{PopIII} \leq 300M_\odot$) and thus emits much more LW radiation (Eq.~\ref{eq:QLW}). 
As shown in the top middle and top right subplot of Fig.~\ref{fig:kdiss_apxc-ray_comparison}, at any given gas density or $\text{H}_{2}$ density, cells that receive less unattenuated LW flux (and thus are further away from the stars) are more likely to be photodissociated at a higher rate when using the approximation model. Since the approximation model assumes an optically thin LW radiation field, any attenuation information between a radiative source and a gas cell is lost outside of a local characteristic length. For example, if there is a dense molecular cloud between a gas cell and a star particle, when we ray trace an LW photon through each cell, that photon will be attenuated during the calculation. On the other hand, the LW radiation in the approximation model can bypass that dense cloud, and the self-shielding characteristic length of the faraway cell cannot capture that information either because the length value is only determined locally. The further a gas cell is from the radiative sources, the fewer LW photons the cell can receive and the more attenuation a light ray can undergo, leading to a greater difference between an optically-thin approximation and a model that takes into account attenuation throughout the ray's long traveled length. In contrast, the higher the LW flux a gas cell receives, the closer it is to the radiative sources, the better the approximation model matches the ray-tracing model because the approximation's characteristic length can better match the attenuation over short distances. \corr{This connection between the unattenuated LW radiation flux and the agreement of the two models can be seen more clearly in the bottom right figure.} \corr{In summary}, the accuracy of $k_\mathrm{diss, apx-c}$ can be positionally and directionally dependent.

We also examine whether the corrected implementation of the Sobolev-like approximation significantly changes the outcomes when compared with the old implementation. We manually calculate the two approximations on the gas cells in the same ``\textit{ray}'' snapshot as in Fig.~\ref{fig:kdiss_apxc-ray_comparison}. Similar to the behavior in the top right subplot of Fig.~\ref{fig:kdiss_apxc-ray_comparison}, Fig.~\ref{fig:kdiss_apxc-apx_comparison} shows that the effect of the correction is most significant for cells with high $\text{H}_{2}$ density because the approximated self-shielding effect is more amplified in these cells. 
%However, after the fix, the $k_\mathrm{diss}$ approximation barely changes. 
Even though the corrected version does increase the photodissociation rate, over 99\% of the volume have the rate change within 10\% when we switch from the old formula to the corrected formula. The maximum change is a factor of 2 increase in $k_\mathrm{diss}$ after applying the correction. Furthermore, gas cells with the greatest change after applying the correction have very small $k_\mathrm{diss,apx}$, and thus the change even has less effect on the $\text{H}_{2}$ on that cell. As we show later in Section~\ref{sec:result}, results from the simulation using the corrected implementation still agree well and remain consistent with results from the simulations using the old implementations. Therefore, all three simulations ``\textit{apx1}'', ``\textit{apx2}'', and ``\textit{apx-c}'' in our analysis can be representatives of the Sobolev-like approximation model, and the correction should have a limited impact on any results run by an older version of ENZO's \texttt{solve\_rate\_cool.F} file.

\subsection{Effects on the \texorpdfstring{$\text{H}_{2}$}{H2} Content of Halos}
\label{subsect:effect_on_H2_content}

\begin{figure}
    \centering
    \includegraphics[width=0.9\columnwidth]{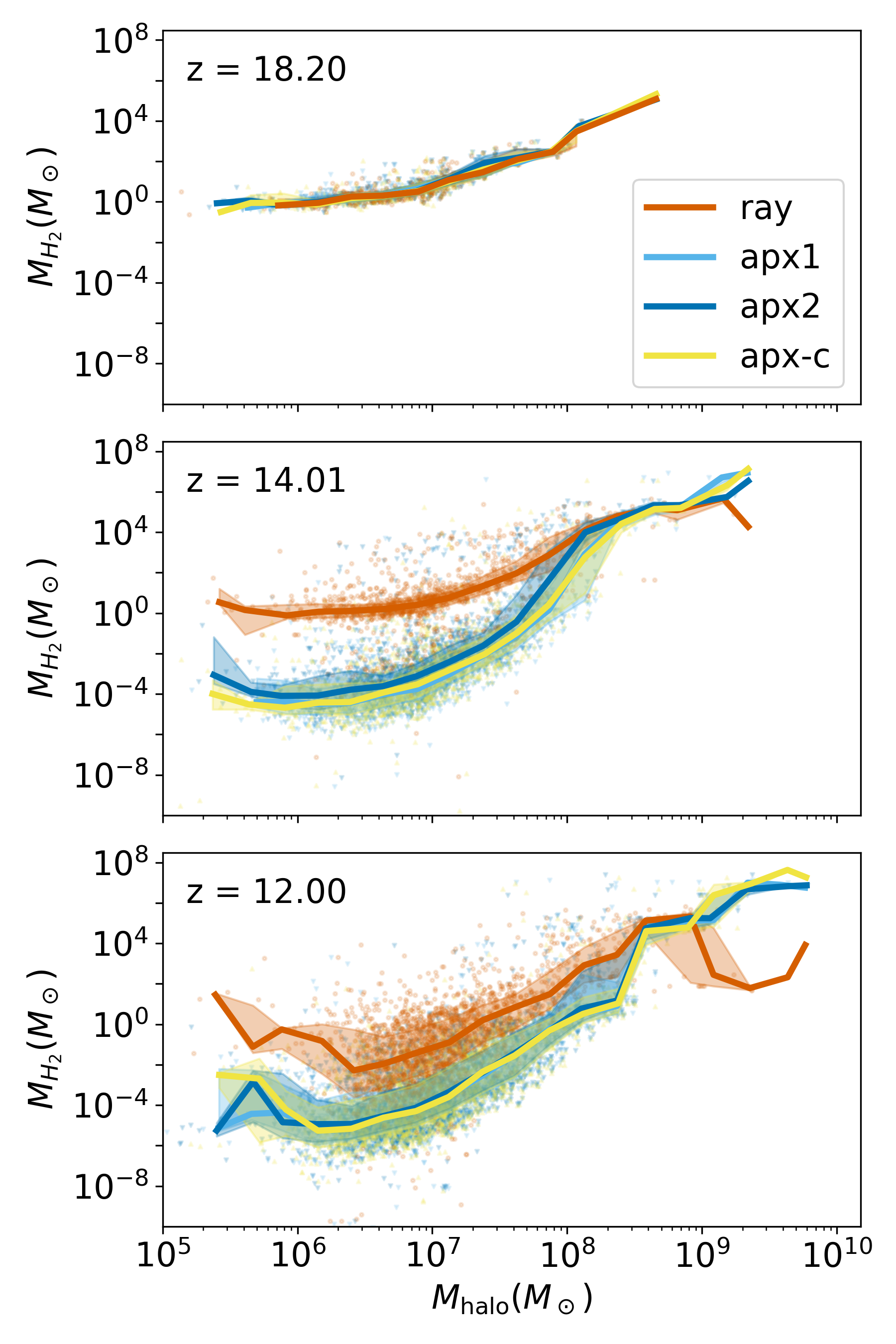}
    \caption{The relationship between $\text{H}_{2}$ mass and halo mass in the timestep where stars first appear, at $z = 14.01$, and in the last timestep of the ``\textit{ray}'' simulation. Each point represents an individual halo. The solid lines represent the median $\text{H}_{2}$ mass for each halo mass bin, and the shaded regions represent the 25th-to-75th percentiles of $\text{H}_{2}$ mass in each bin. The halo's $\text{H}_{2}$ mass differs substantially between the two models, especially for small halos ($M_\text{halo} \leq 10^{8} M_\odot$) and massive halos ($M_\text{halo} \geq 10^{9} M_\odot$).}
    \label{fig:H2_frac}
\end{figure}

We proceed to investigate the amount of $\text{H}_{2}$ in halos of the ray-tracing and approximation treatments. Fig.~\ref{fig:H2_frac} shows the $\text{H}_{2}$ mass as a function of the halo mass at three redshifts. The first redshift is at the time where star particles first appear in the simulations (z $\approx$ 18.20), the second is at a time step in the middle of the simulation runs (z $\approx$ 14.01), and the third one is the last time step of the ``\textit{ray}'' run (z $\approx$ 12.00). Each halo in the snapshot is represented by a scatter point. The solid lines and the shaded regions represent the median values and the interquartile range of the molecular hydrogen mass ($M_{\text{H}_{2}}$) for each halo mass bin. Despite random mass sampling from the IMF during PopIII star formation, all three simulations ``\textit{apx1}'', ``\textit{apx2}'', and ``\textit{apx-c}'' have a comparable distribution of halos in the $M_{\text{H}_{2}}\text{--}M_\text{halo}$ space, showing that the different realizations of the approximation models still converge in the same trend. On the other hand, the molecular gas mass shows a notable difference between the two radiative transfer treatments, especially after stars form and in small halos ($M_\text{halo} < 10^{8}M_\odot$) and in large halos ($M_\text{halo} > 10^{9}M_\odot$). At the last time step of our analysis, for halos whose halo mass is smaller than $10^{8} M_\odot$, $M_{\text{H}_{2}}$ in the ray-tracing treatment is larger than that of the Sobolev-like approximation treatment by two to three orders of magnitude. Indeed, the median $M_{\text{H}_{2}}$ for halos at this mass range is $10^{-2}\text{--}10^{1} M_\odot$ in the ``\textit{ray}'' simulation and about $10^{-4}\text{--}10^{-1} M_\odot$ in the \textit{apx} simulations. The 25th-to-75th percentile range of the $M_{\text{H}_{2}}$ distribution does not even overlap between the two radiative models, proving that most of the halos in the ray-traced simulation have more molecular gas than the halos in the simulations with the approximation. This shows that the self-shielding effect is much stronger in smaller halos when using the ray-tracing method, and thus it helps preserve more $\text{H}_{2}$ in the halos. This is consistent with Fig.~\ref{fig:kdiss_apxc-ray_comparison}, where we observe that the approximation treatment considerably boosts the $\text{H}_{2}$ photodissociation rate in a low gas density regime, inferring that smaller halos are the most susceptible to the choice of the model. 

On the other hand, the effect is opposite for halos with halo mass over $10^{9}M_\odot$. The $\text{H}_{2}$ mass in these halos is about five orders of magnitude lower in the ``\textit{ray}'' simulation than in any of the \textit{apx} simulations. This is a consequence of the over-shielding effect at high gas density when using the density-gradient approximation treatment (top middle subplot of Fig.~\ref{fig:kdiss_apxc-ray_comparison}). Because of their deep potential wells, massive halos can accumulate more gas and retain gas better, allowing gas to reach higher density. Denser gas results in faster cooling. This causes gas to collapse more easily, and therefore, stars are more likely to form in these halos. As a result, the feedback from these stars photodissociates the surrounding molecular clouds and decreases the halo's $\text{H}_{2}$ content. However, when using the approximation treatment, these molecular clouds receive enhanced shielding and the photodissociation rate is strongly suppressed. Therefore, the $\text{H}_{2}$ fraction remains high in the halos in the \textit{apx} simulations. 

We see that for large halos, the discrepancy grows greater over time because gas reaches to a higher density in these halos as they grow larger, leading to more overestimation of self-shielding when using the approximation. On the other hand, for small halos, the discrepancy in halos' molecular gas content peaks at $z \approx 14\text{--}15$ and then decreases. When stars first show up in the \textit{apx} simulations, because the model assumes an optically-thin LW field, LW radiation can freely travel to all halos and dissociate $\text{H}_{2}$ there. In contrast, the LW radiation from the first stars in the ray-traced simulation is attenuated on its way out, and thus it only destroys $\text{H}_{2}$ near the radiative sources. As there are more stars being formed and more LW radiation being emitted throughout the box, smaller halos in the ``\textit{ray}'' simulation that are further away from radiative sources start to be affected, alleviating the discrepancy in the halos' $\text{H}_{2}$ mass between the two models. This effect can be seen more clearly when we investigate the larger simulated volume in Subsection~\ref{subsect:larger_volume}.

In summary, Fig.~\ref{fig:H2_frac} displays a clear shift of $M_{\text{H}_{2}}$ difference between the ``\textit{ray}'' and the \textit{apx} simulations when going from low-mass to high-mass halos. When using the Sobolev-like approximation, small halos lose $\text{H}_{2}$ easily while massive halos over-conserve $\text{H}_{2}$. The halo mass that the approximation agrees the best with the ray-tracing method is about $5\times10^{8}$ $M_\odot$.  

\subsection{Effects on Star Formation}
\label{subsect:effect_on_sf}

\begin{figure*}
	\gridline{\fig{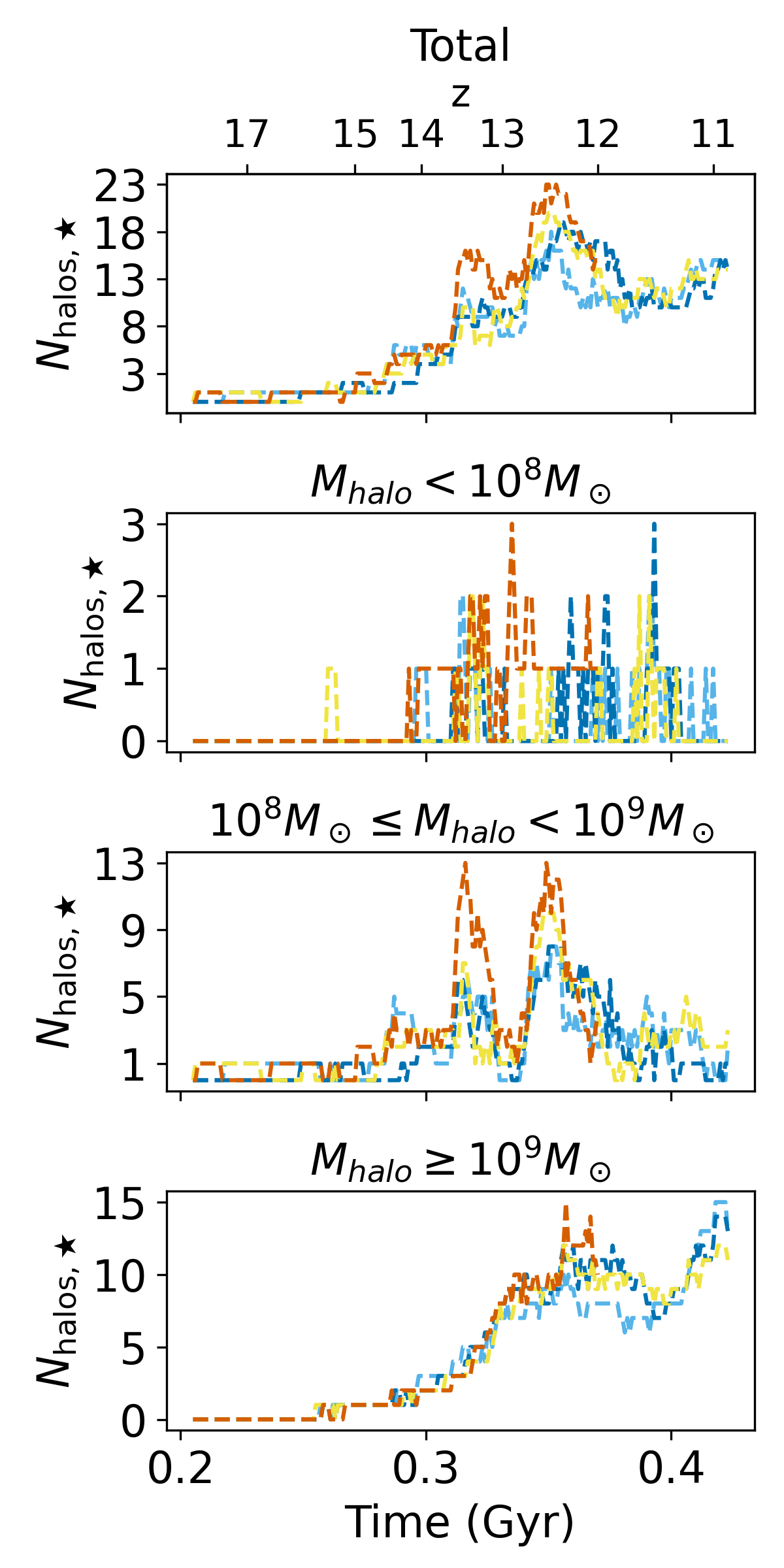}{0.335\textwidth}{(a)}
    \fig{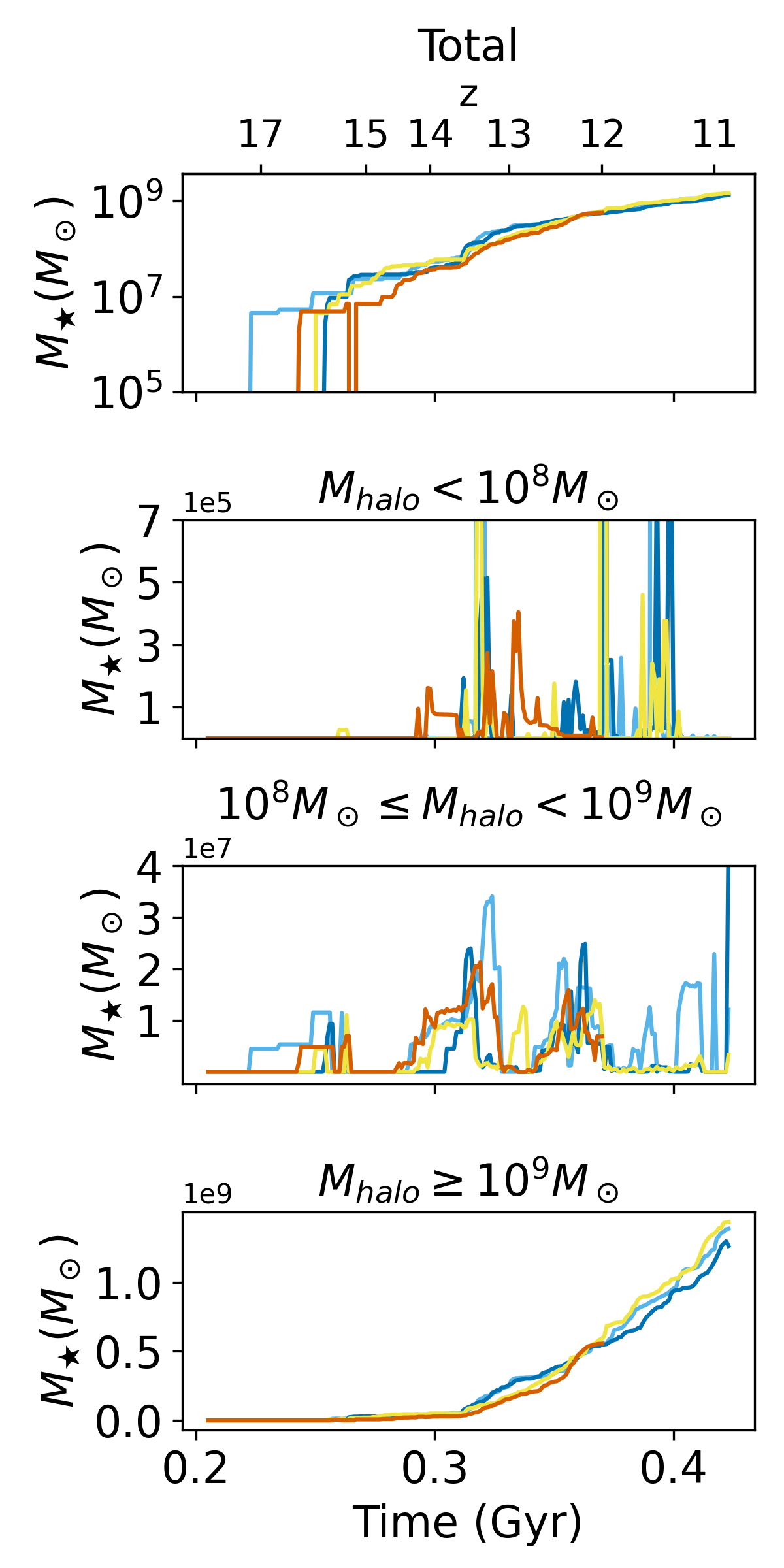}{0.335\textwidth}{(b)}
	\fig{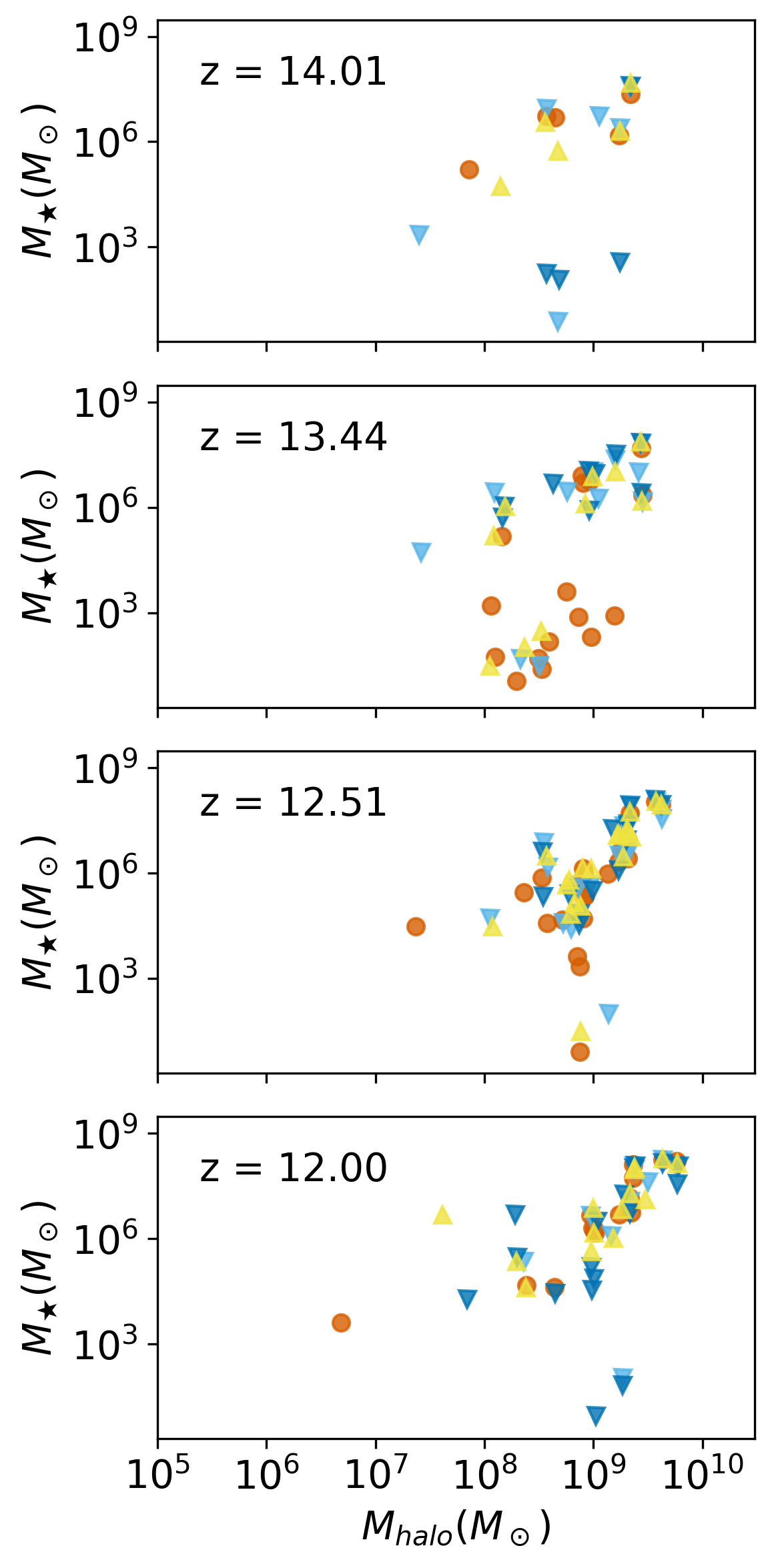}{0.3\textwidth}{(c)}}
	\caption{Column (a): The comparison of the number of halos with stars as a function of time for each halo mass bin between our four simulations. Column (b): The comparison of the total bounded stellar mass as a function of time for each halo mass bin. Column (c): The relationship between stellar mass and halo mass of all the halos with stars at different redshifts. Up to $z = 12$, there are more small halos with stars in the ``\textit{ray}'' simulation. However, the number of larger halos with stars is more consistent between the simulations.}
	\label{fig:stars}
\end{figure*}

We also investigated whether the self-shielding treatment affects the stellar mass of galaxies. Column (a) of Fig.~\ref{fig:stars} displays the number of halos that host stars (which we refer to as starry halos) as a function of time for different halo dark matter mass bins. The mass bins are chosen to reflect the varying accuracy of the approximation model in predicting the halo's $\text{H}_{2}$ mass (Fig.~\ref{fig:H2_frac}). The stellar component of the halos is calculated by using the star assignment procedure described in Section~\ref{subsect:star_assignment}. Each star is assigned to a unique halo, and therefore, there is no double counting in stellar mass between halos. From the top plot of Column (a), we can see the timing of the first star formation in the simulation is also not correlated with the choice of the radiative transfer model, as the \textit{apx} simulations can form stars both before and after the ``\textit{ray}'' simulation. Column (b) shows the total bounded stellar mass of all halos in each halo mass bin. Column (c) shows the stellar mass-halo mass relationship at different redshifts. The strong dissimilarity in the $M_{\text{H}_{2}}\text{--} M_\mathrm{halo}$ relation between the two $\text{H}_{2}$ radiative transfer models found in Fig.~\ref{fig:H2_frac} is reflected in the halo's stellar mass when we notice that there are more small halos forming stars in the ``\textit{ray}'' simulation than in any of the realizations of the \textit{apx} simulations. This effect is most noticeable for halos with dark matter mass between $10^{8} M_\odot$ and $10^{9} M_\odot$, as shown in the third row of Column (a). For smaller halos ($M_\text{halo} < 10^{8} M_\odot$), the gravitational well may not be deep enough for stars to form effectively until the halos get to a larger dark matter mass, and thus we do not see a lot of halos with stars or a lot of stellar mass formed in this halo mass bin (second row of Columns (a) and (b)). For halos with mass $10^{8} M_\odot \leq M_\text{halo} < 10^{9} M_\odot$, even though they form stars more easily in the ``\textit{ray}'' simulation, they do not form a lot of stars, and the stellar mass each halo form is quite small ($M_\star < 10^{3} M_\odot$, as shown in the second and third rows of Column (c)). Indeed, the third row of Column (b) displays that, compared to the \textit{apx} simulations, the total stellar mass created by halos in this mass bin is comparable or even slightly smaller in the ``ray'' simulation. Even though there are fewer starry halos in the \textit{apx} simulations, each of these small starry halos has more stellar mass than the \textit{ray} counterparts because the approximation model starts to over-shield $\text{H}_{2}$ at intermediate gas density (top middle plot of Fig.~\ref{fig:kdiss_apxc-ray_comparison}). Also, many small starry halos in the ``\textit{ray}'' contain mostly PopIII stars. For example, at $z = 13.44$ (second row of Column (c)), 9 out of 16 \textit{ray}'s starry halos contain only PopIII stars. Given that PopIII stars can only form in $\text{H}_{2}$-rich gas (we set the $\text{H}_{2}$ fraction threshold to be 0.001 before a PopIII star formation is considered), this is more possible when using the ray-tracing models because $\text{H}_{2}$ is protected better in low gas density.
%Also, because we set the minimum mass of our PopII star cluster particle to be $10^{3} M_\odot$, when many small starry halos in the ``\textit{ray}'' simulation have a stellar mass of $\approx 10^{3} M_\odot$, this suggests that these halos contain mostly PopIII stars. For example, at $z = 13.44$ (second row of Column (c)), 9 out of 16 \textit{ray}'s starry halos contain only PopIII stars. 
%This is reasonable because the allowed mass of our PopIII star particle is smaller than the allowed mass of our PopII star particle by at least half an order of magnitude, meaning that for small halos with a limited amount of molecular gas, PopIII stars form more easily. This is more possible when using the ray-tracing models because $\text{H}_{2}$ is protected better in low gas density.

\corr{It is important to note that our simulations' mass resolution and refinement strategy may also play a factor in hindering star formation in small and mini halos. Previous studies of PopIII stars suggested that the minimum mass of star-forming halos could be between $10^{5}$ and $10^{6} M_\odot$ \citep{Schauer+2019, CorreaMagnus+2024, Lenoble+2024}. The dark matter resolution in these works ranges from $1$ to $10^{3} M_\odot$, which is more capable of resolving minihalos and following the production of cold gas in minihalos than in our simulations. Given our simulation's dark matter mass resolution, our halos are relatively well-defined down to $\approx 2\times10^{6} M_\odot$ (75 particles), and we do find stars in some halos around this mass. Below $\approx 2\times10^{6} M_\odot$, the halos become less defined, which could affect the formation of cold gas and stars.  Thus, the absence of stars in our small halos ($\leq 2\times10^{6} M_\odot$) in all four simulations could be due to the mass resolution rather than the $\text{H}_{2}$ radiative treatment, which makes it challenging to evaluate the treatment's effect in minihalos.}

On the other hand, for larger halos ($M_\text{halo} > 10^{9} M_\odot$), the number of halos with stars is consistent between the simulations. The total stellar mass in large halos is slightly lower when using the ray-tracing model (by 20 to 30 percent as a median value of all snapshots, as shown by the fourth row of Column (b)), which can be explained by a lower amount of molecular gas in the ``\textit{ray}'' halos (bottom plot of Fig.~\ref{fig:H2_frac}). Interestingly, the $\text{H}_{2}$ radiative transfer treatments affect photodissociation rates and $\text{H}_{2}$ content considerably, but star formation in these halos is not strongly impacted, at least to $z = 12$. Hence, we investigated the star-forming environments and star formation criteria to explain why the strong differences in photodisociation rates do not appear to substantially affect the stellar mass of larger halos. In our simulations, we have set the overdensity threshold (with respect to the \corr{total} mean matter density) for the formation of both PopIII star particles and PopII star cluster particles to be $10^{6}$. As the mean matter density of the universe is a function of redshift, these overdensities correspond to the physical density of \corr{$\approx 2\times10^{-20} \text{g}/\text{cm}^{3}$} at $z = 18.2$ (which is the first snapshot that has stars) and decreases to the physical density of $\approx 6\times10^{-21} \text{g}/\text{cm}^{3}$ at $z = 12$. 
%In our simulations, we have set the overdensity threshold for the formation of PopIII star particles to be $10^{6}$. For PopII star cluster particles, the overdensity threshold is set to $10^{7}$ \citep{Wise+2009}. As the critical density of the universe is a function of redshift, these overdensities correspond to the physical density of $\approx 0.2\text{--}2\times10^{-19} \text{g}/\text{cm}^{3}$ at $z = 18.2$ (which is the first snapshot that has stars), which decreases to the physical density of $\approx 0.6\text{--}6\times10^{-20} \text{g}/\text{cm}^{3}$ at $z = 12$. 
According to Fig.~\ref{fig:kdiss_apxc-ray_comparison}, a gas density between \corr{$\approx 10^{-21}\text{--}10^{-20} \text{g}/\text{cm}^{3}$} corresponds to the range where the Sobolev-like approximation of the photodissociation rate more likely matches the ray-tracing calculation. In other words, for a cell that just reaches an overdensity sufficient for star formation, its molecular content can be similar between the two radiative treatments, so a comparable amount of stellar mass will be formed in these cells. \corr{As shown later in Subsection~\ref{subsect:case_study}, another factor contributing to the convergence of stellar mass for large halos between methods is metal cooling. After producing a lot of stars, these halos will be enriched by metals from stellar feedback, which then contribute to cooling in addition to $\text{H}_{2}$ cooling, allowing star formation to happen despite the strong difference in the $\text{H}_{2}$ amount between the two models.}

It should be noted that this conclusion has only been examined up to $z = 12$ and can be further explored in future work. At lower redshifts \corr{where the fixed overdensity threshold criterion for star formation is still appropriate}, the mean matter density of the universe decreases, and thus the gas density \corr{threshold} for star formation also decreases. \corr{This density may lie in a regime where the approximation treatment starts to overpredict the photodissociation rate, resulting in a lower amount of $\text{H}_{2}$, a longer cooling time, and consequently a harder condition for stars to form. Consequently, smaller halos and dwarf galaxies that do not have a very deep potential well to reach high gas density may be inhibited from forming stars when using the Sobolev-like approximation. On the contrary, gas-phase metallicity can be enriched at lower redshifts after multiple generations of stars, allowing gas to cool via the cooling channels of metal and molecules other than $\text{H}_{2}$ (for example, CO cooling,  \cite{Glover+2012}). Thus, depending on the metallicity of the halos, the influence of the $\text{H}_{2}$ radiative transfer models on star formation may become less impactful at lower redshift and requires further investigation.}

%For example, at $z = 0$, the Universe's mean matter density is $\approx 3\times10^{-30} \text{g}/\text{cm}^{3}$. With an overdensity threshold of $10^{6}$ for star formation, this requires a gas density of $\approx 3\times10^{-24} \text{g}/\text{cm}^{3}$. This density lies in a regime where the approximation treatment starts to overpredict the photodissociation rate, resulting in a lower amount of $\text{H}_{2}$, a longer cooling time, and consequently a harder condition for stars to form. Therefore, we expect that if we run the simulations to the present time, the effect of the models on star formation will be much more noticeable. Smaller halos and dwarf galaxies, especially isolated ones that do not have a very deep potential well to reach high gas density, will be inhibited from forming stars when using the Sobolev-like approximation for $\text{H}_{2}$ radiative treatment.

\subsection{Individual Halo Case Study}
\label{subsect:case_study}

\begin{figure*}
	\gridline{\fig{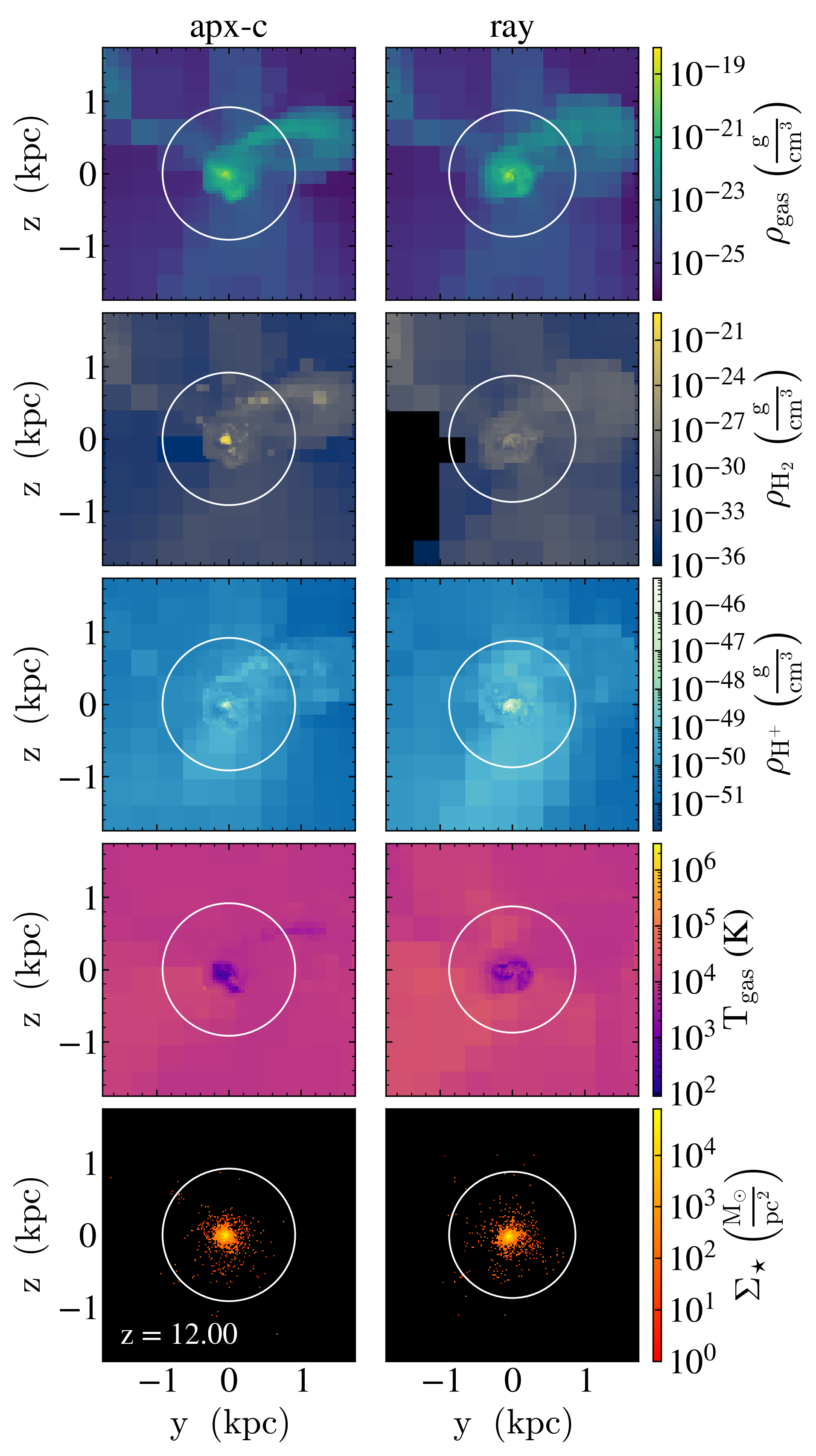}{0.375\textwidth}{(a)}
		\fig{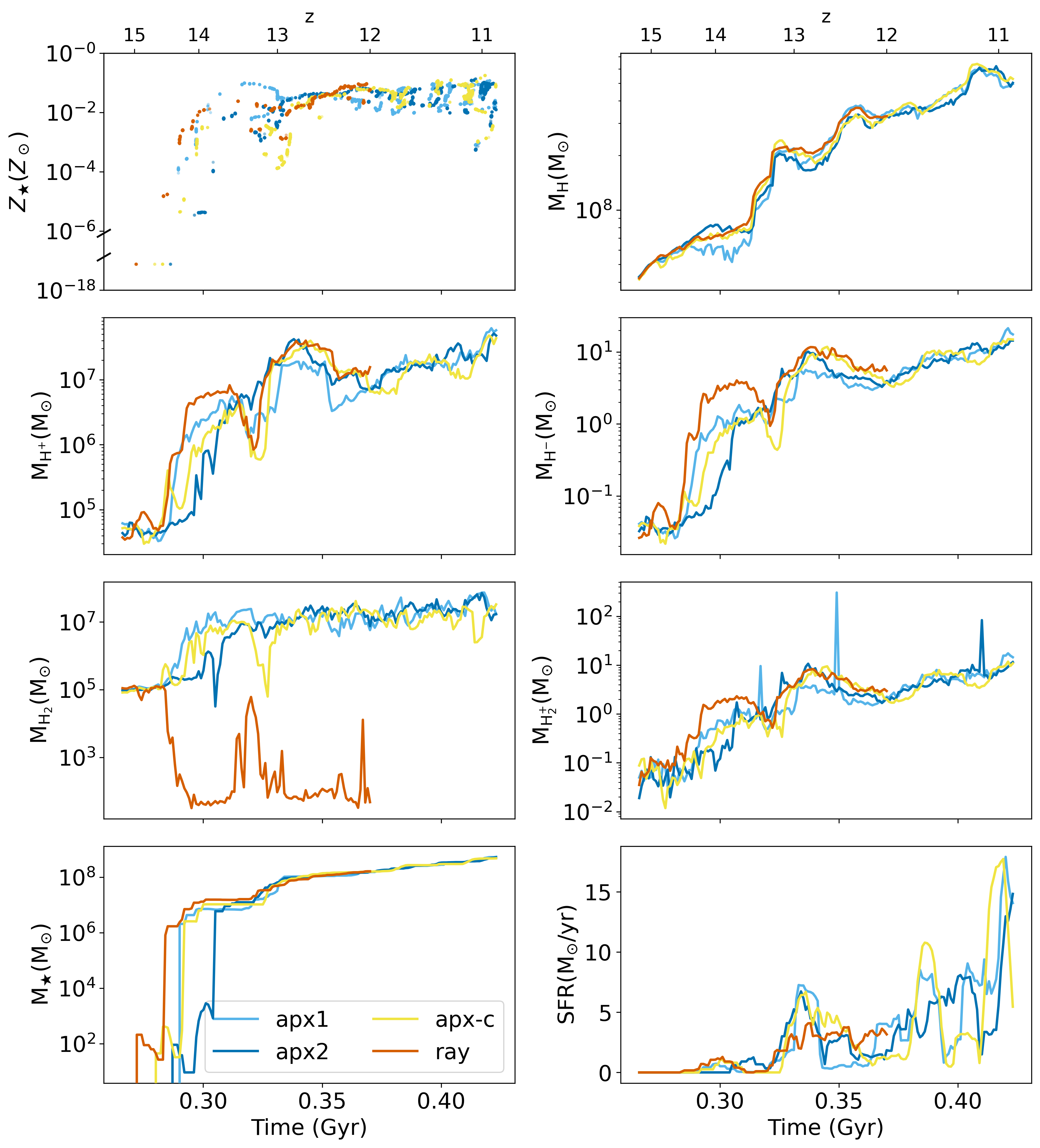}{0.615\textwidth}{(b)}}
	\caption{Subplots (a): The gas density, $\text{H}_{2}$ density, \corr{$\text{H}^{+}$} density, temperature, and stellar surface density projection plot of the largest halo in our simulations ($M_{\mathrm{halo}} = 6\times10^{9} M_\odot$ at $z$ = 12). The white circle represents the galaxy's boundary, which is defined as the radius enclosing the baryonic (star + gas) density 2000 times larger than the universe's critical density. Subplots (b): The time evolution of \corr{stellar metallicity,} different hydrogen species' mass, the stellar mass, and the star formation rate between the simulations. Among these properties, the amount of $\text{H}_{2}$ and \corr{$\text{H}^{+}$} differ the most between the two models, especially near the galactic center.}
	\label{fig:Halo0-0_comparison}
\end{figure*}

\begin{figure*}
    \gridline{\fig{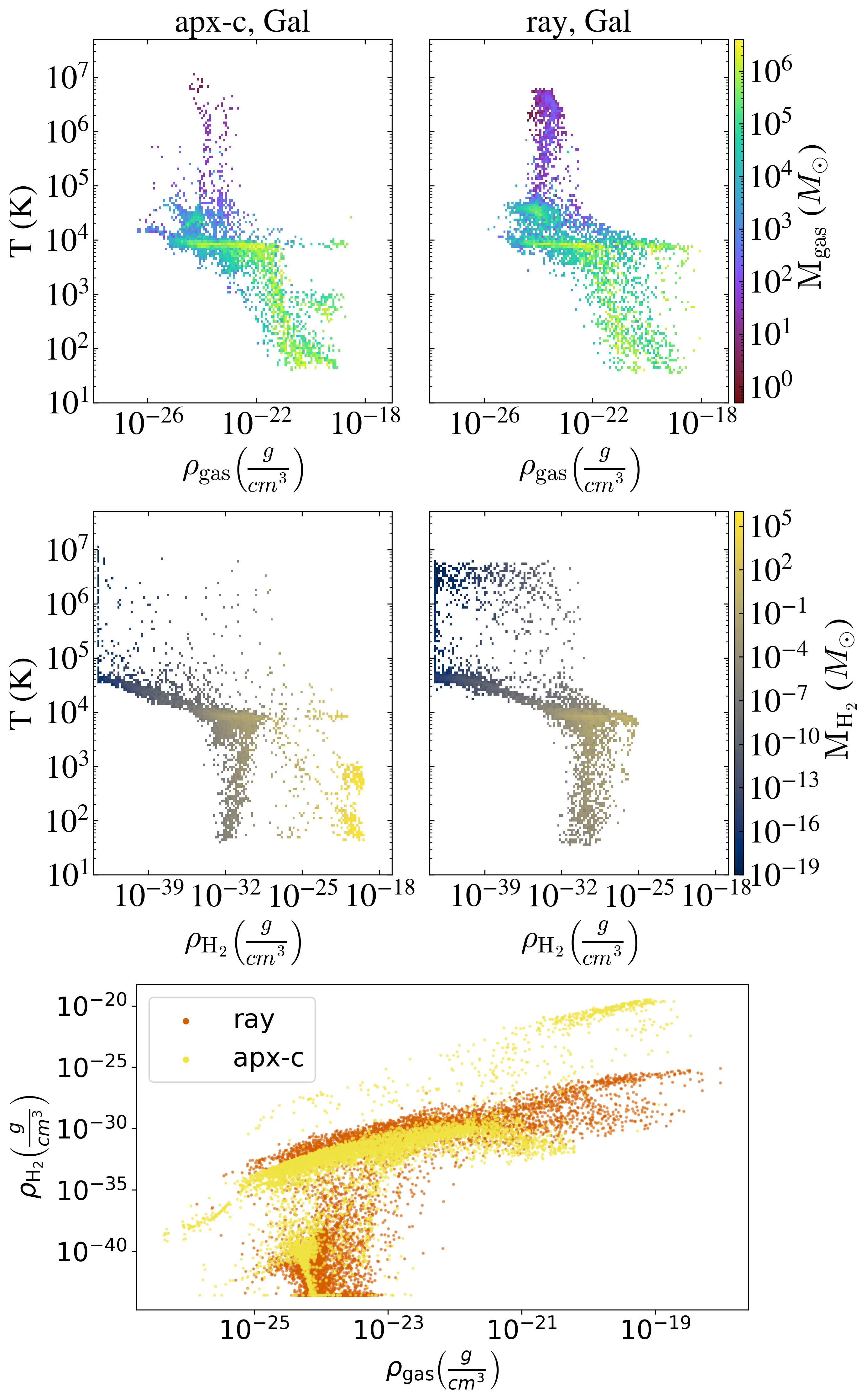}{0.48\textwidth}{(a)}
    \fig{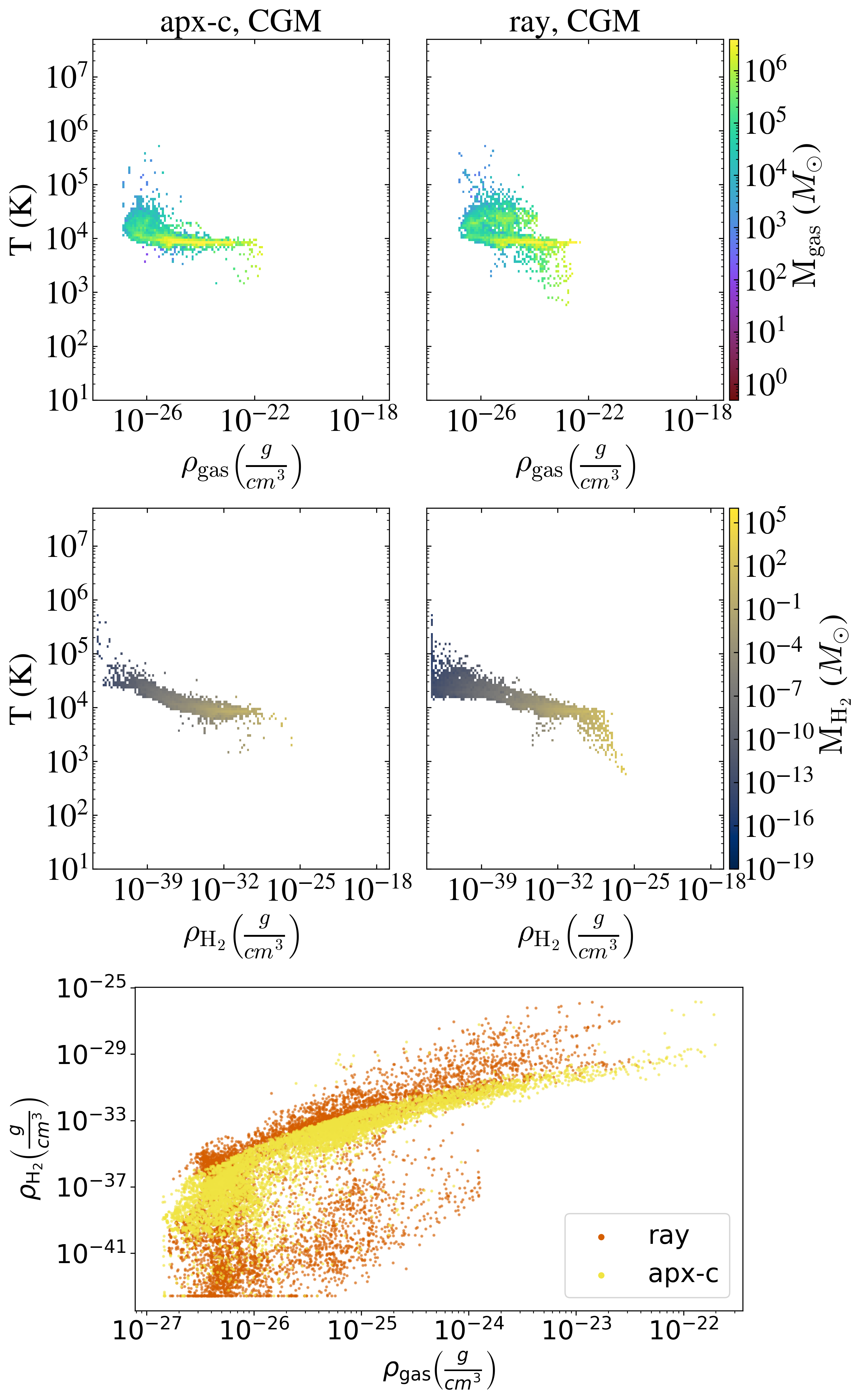}{0.48\textwidth}{(b)}}
    \caption{Gas and $\text{H}_{2}$ phase plots colored by their mass in the galaxy ($r \leq R_{2000,\text{bary}}$, Subplots (a)) and in the CGM ($R_{2000,\text{bary}} \leq r \leq 5R_{2000,\text{bary}}$, Subplots (b)) of the largest halo in our simulation set, evaluated at $z = 12$. For each subplot, the left column shows the halo in the ``\textit{apx-c}'' run, and the right column shows its counterpart in the ``\textit{ray}'' run. The bottom row shows the relationship between the $\text{H}_{2}$ density and the gas density of all gas cells in the halo. The use of the self-shielding model affects the temperature, gas density, and $\text{H}_{2}$ density profiles of both the galaxy and the CGM.  } 
    \label{fig:phaseplot_Halo0-0}
\end{figure*}

To better examine the effect of the $\text{H}_{2}$ radiative transfer on each halo, we chose one halo to perform a case study. By $z = 12$, we have seven isolated main halos that can be cross-matched between the four simulations using the procedure in Subsection~\ref{subsect:cross-matching_halos}. However, none of our simulations have small starry halos that are isolated, as shown later in Fig.~\ref{fig:starposition_allbox}. All small halos with stars in our simulations are sub-halos forming stars during mergers or stealing star particles from the progenitor halo during pericentric passage. Because a galaxy merger is a dynamically complex and intricate process whose results are extremely sensitive to the halo properties, it is difficult to study the evolution of small halos between the four simulations. Thus, we only perform a case study on large and isolated halos in our simulated box. 

We chose the largest halo in our four simulations for this case study because it forms stars early, and hence we can see how the two $\text{H}_{2}$ models affect a halo for a longer period of time. Other large isolated halos show relatively similar behaviors and will be discussed in Subsection~\ref{subsect:radialprofile}.  
%which serves as an example of how a halo evolves when using two types of $\text{H}_{2}$ radiative transfer models. 
In Subplots (a) of Fig.~\ref{fig:Halo0-0_comparison}, we show the projection of various properties at the last time step ($z = 12$) between ``\textit{apx-c}'' (left column) and ``\textit{ray}'' (right column). Subplots (b) of the figure display the time evolution of all hydrogen species mass within the galaxy's ISM region ($R < R_{2000,\text{bary}}$), \corr{metallicity,} the halo's stellar mass, and the halo's SFR. Because the total reservoir of protons is held constant in our 9-species chemistry model, \corr{the total gas mass stays consistent between our simulations throughout the halo's evolution and is unaffected by the choice of $\text{H}_{2}$ self-shielding model}. The gas density projection plot also shows a similar galaxy morphology between our runs. However, even though there is an agreement on total gas content, the amount of $\text{H}_{2}$ is considerably different between the two models. The mass of $\text{H}_{2}$ starts to diverge around $z \approx 14.5$ when stars start to form in the ``\textit{ray}'' simulation, and by $z = 12$, the ``\textit{ray}'' galaxy has about $10^{-5}$ times as much $\text{H}_{2}$ as the galaxy in any of the other \textit{apx} simulations. This stark dissimilarity in $\text{H}_{2}$ mass is also reflected in the $\text{H}_{2}$ projection plot (second row of Subplots (a)), where we see a discernible clump of molecular gas in the center of the ``\textit{apx-c}'' galaxy but not in the ``\textit{ray}'' galaxy. Due to radiative cooling, the higher concentration of $\text{H}_{2}$ lowers the temperature of the galactic center, as shown in the temperature projection plot. The greatest difference in $\text{H}_{2}$ between the models is focused in the very center region of the galaxy, where a lot of LW radiation is produced and gas density is high ($\rho_\text{gas} \approx 10^{-18} \text{g}/\text{cm}^{3}$), leading to the approximation overestimating the amount of shielding more substantially (Fig.~\ref{fig:kdiss_apxc-ray_comparison}). Indeed, the discrepancy of $\text{H}_{2}$ in the galactic center likely drives the discrepancy in the $\text{H}_{2}$ mass of massive halos that we see in Fig.~\ref{fig:H2_frac}. The ``\textit{apx-c}'' galaxy is also more clumpy overall as it other smaller clumps of molecular gas outside of the galactic center. For these smaller clumps, even though the gas density outside of the galactic center is lower ($\rho_\text{gas} \approx 10^{-20} \text{g}/\text{cm}^{3}$), these cells are still located near the active young star particles and receive more LW flux. Thus, the photodissociation rate is still likely to be \corr{lower} when using the approximation treatment, a result shown in Fig.~\ref{fig:kdiss_apxc-ray_comparison}. In contrast, the halo in the ``\textit{ray}'' simulation displays a more uniform distribution of molecular gas.

When comparing the halo's stellar mass, we do not see significant differences. Even though there is a timing mismatch of when stars first form in our four simulations, the halos' stellar mass converges after $z \approx 13$. At $z = 12$, we also see convergence in stellar mass within half an order of magnitude in the other six isolated main halos, though the stellar mass is slightly lower in the ``\textit{ray}'' simulation, as remarked in Subsection~\ref{subsect:effect_on_sf}. \corr{The top left figure of Subplots (b) shows the metallicity of stars when they form to highlight the exact contribution of gas metallicity to each star's formation. At very high redshift ($z > 14$), stars form in metal-poor gas cells ($Z < 10^{-4} Z_\odot)$, where $\text{H}_{2}$ cooling plays a main role. Later generations of stars form in a more metal-rich gas cells ($Z \approx 0.1\text{--}0.2 Z_\odot$) as the simulation evolves. This enrichment in metals allows more channels for the gas to cool down to hundreds of Kelvin in addition to the $\text{H}_{2}$ cooling \citep{Smith+2008}, bridging the discrepancy in stellar mass.} Despite that, we noticed a slight discrepancy in the radial stellar mass distribution, which will be explored later in Subsection~\ref{subsect:radialprofile}. In terms of SFR, we do not see the same burst of star formation in the ``\textit{ray}'' galaxy as in the \textit{apx} galaxies. However, we do not notice a consistent enough trend in other halos' SFR to draw a strong conclusion about whether the $\text{H}_{2}$ radiative model changes the star formation history. 

Because the galaxy's total gas mass and the stellar mass is relatively similar across simulations, the decline of $\text{H}_{2}$ in the ``\textit{ray}'' data is largely explained by the conversion of $\text{H}_{2}$ into other hydrogen species rather than by $\text{H}_{2}$ being used for star formation. According to the second and third rows of Subplots (b) of Fig.~\ref{fig:Halo0-0_comparison}, when the $\text{H}_{2}$ mass starts to drop in the ray-tracing model, \corr{$\text{H}_{2}^{+}$}, \corr{$\text{H}^{+}$}, and $\text{H}^{-}$ mass compensates. Among those three species, the increase in \corr{$\text{H}^{+}$} mass is the largest and helps compensate for most of the decrease in $\text{H}_{2}$ mass. We can also observe this in the \corr{$\text{H}^{+}$} projection plot (third row of Subplots (a)) with the ``\textit{ray}'' galaxy having a much higher concentration of \corr{$\text{H}^{+}$} near the galactic center and also a higher amount of \corr{$\text{H}^{+}$} throughout the ISM. In other words, the LW radiation from nearby stars photodissociates the molecular gas to atomic hydrogen, and then ionizing radiation immediately ionizes the atomic hydrogen to become ionized hydrogen. We do not see a simultaneous increase in \corr{H} mass as $\text{H}_{2}$ mass decreases because after $\text{H}_{2}$ is photodissociated, the resulting \corr{H} is very quickly ionized. It is important to note that conversions between hydrogen species occur in all four simulations. We can also see the fall of $\text{H}_{2}$ mass and the corresponding increase in other hydrogen species in the \textit{apx} simulation. We can also notice that \corr{$\text{H}_{2}^{+}$}, \corr{H}, \corr{$\text{H}^{+}$}, and $\text{H}^{-}$ react with each other to replenish $\text{H}_{2}$ in the galaxy \citep{Galli+1998}, which is most noticeable at $\approx$ 0.32 Gyr in the ``\textit{ray}'' galaxy. The main distinction of the \textit{apx} simulations is that  $\text{H}_{2}$ becomes so self-shielded that it is much harder to be destroyed than in the ``\textit{ray}'' simulation, leading to a higher amount of $\text{H}_{2}$ in the galaxies in the these simulations. 

With respect to the two realizations of the universe that use the old implementation of the approximation models (``\textit{apx1}'' and ``\textit{apx2}''), we notice certain differences. Throughout the evolution of the halo, the ratio of $M_{\text{H}_{2}}$, \corr{$M_{\text{H}^{+}}$}, and $M_{\bigstar}$ between ``\textit{apx1}'' and ``\textit{apx2}'' are $1.13^{+1.1}_{-0.31}$, $0.99^{+0.23}_{-0.40}$, $0.99^{+0.22}_{-0.25}$, respectively, where the lower and upper bounds represent the 25th to 75th quartile range. This shows that due to the stochastic mass sampling of PopIII star formation, different realizations of the same parameters file usually change the halo's gas mass and the stellar mass no more than a factor of two at any given time. This further emphasizes the great discrepancy of $M_{\text{H}_{2}}$ when using the ray-tracing model because $M_{\text{H}_{2}}$ in the ``\textit{ray}'' simulation is lower by at least two orders of magnitude compared to any of the \textit{apx} simulations. Furthermore, we can also see that the properties of ``\textit{apx-c}'' simulation are also consistent with those of ``\textit{apx1}'' and ``\textit{apx2}'', proving that the correction in the implementation does not alter the general results.

%once again assert that despite having a slight discrepancy in star formation history, the properties of their halo still converge within half order of magnitude after at most 50 million years since the halo starts having stars. Therefore, the stochastic mass sampling of PopIII star formation should not considerably change the simulation outcomes. 

%Furthermore, throughout most of this halo's evolution, the stellar mass in \textit{EarlyRe-ray} stays about several dex higher in the \textit{EarlyRe-apx} version and relatively converge at the last snapshot. Thus, this shows that the choice of $\text{H}_{2}$ self-shielding model affects the timing of star formation in halos while the stellar mass between two models will eventually converge. While the ray-tracing model might initially boost star formation due to higher amount of $\text{H}_{2}$, this effect will be counteracted by increased stellar feedback later, leading to a balanced average on stellar mass.  

We examined the ISM and CGM region of the halo in this case study to further understand the effect of $\text{H}_{2}$ self-shielding models in different parts of a halo. Fig.~\ref{fig:phaseplot_Halo0-0} shows the phase plots of gas and $\text{H}_{2}$ of the halo at $z = 12$. The plots are separated into the ISM region (Subplots (a)) and the CGM region (Subplots (b)). As shown in the projection plots, one noticeable discrepancy is the existence of a large amount of cold dense molecular gas ($T < 10^3 \text{K},\,\rho_{\text{H}_{2}} \approx 10^{-20} \text{g}/\text{cm}^{3}$) only in the ``\textit{apx-c}'' run. The existence of more $\text{H}_{2}$ lowers the gas temperature of the ``\textit{apx-c}'' galaxy and creates cold gas cells. Also, in the gas phase plot, we can see a high concentration of gas at $10^{4} \text{K}$ in both models. This is because with \corr{H} being the gas's dominant component, processes such as Bremsstrahlung, collisional ionization, and collisional excitation cool the gas down to $10^{4} \text{K}$, leading to a large amount of gas at this temperature. In regions where $\text{H}_{2}$ exists, it introduces more sensitivity to energy transitions thanks to the addition of the rovibrational modes and the $\text{H}_{2}$'s electronic energy levels, hence providing more pathways to radiative cooling and further lowering the gas temperature. In the ``\textit{apx-c}'' simulation, the gas in the galaxy cools down after reaching above a density of $\approx 5\times10^{-21} \mathrm{g}/\mathrm{cm}^{3}$ because the approximation model underpredicts $k_\mathrm{diss}$ and allows $\text{H}_{2}$ to exist at large quantity in these dense cells. Indeed, looking at the bottom row of Subplots (a) of Fig.~\ref{fig:phaseplot_Halo0-0}, we see that in the ``\textit{apx-c}'' simulation, $\text{H}_{2}$ exists most abundantly in dense gas cells. There exists an abrupt jump in the $\rho_{\text{H}_{2}}-\rho_\text{gas}$ relationship for the ``\textit{apx-c}'' simulation, while the ``\textit{ray}'' simulation displays a more smoothly increasing trend. Compared to the ray-tracing method, molecular gas in the \textit{apx-c} run is additionally protected in gas cells with $\rho_{\mathrm{H}_{2}} > 5\times10^{-21} \mathrm{g}/\mathrm{cm}^{3}$ and additionally suppressed in gas cells with $\rho_{\mathrm{H}_{2}} < 10^{-21} \mathrm{g}/\mathrm{cm}^{3}$. This plot paints a similar picture as Fig.~\ref{fig:kdiss_apxc-ray_comparison}. In other words, these dense $\text{H}_{2}$ cells are a result of a numerical approximation rather than the actual phenomenon as predicted by the more accurate ray-tracing method. It is important to note that even though there is less cold gas in the ``\textit{ray}'' galaxy and more dense gas at $10^{4}\text{K}$, gas can still cool down to below $100K$ and reach a dense enough density to form stars.

In the CGM, gas has a lower density and thus the $\text{H}_{2}$ photodissociation rate by LW radiation is stronger when using the approximation model (top middle subplot of Fig.~\ref{fig:kdiss_apxc-ray_comparison}). Indeed, the CGM of the galaxy in the ``\textit{ray}'' simulation has more $\text{H}_{2}$ and therefore is slightly colder than the galaxy in the ``\textit{apx-c}'' simulation. Some gas cells in the ``\textit{ray}'' galaxy can be even colder than $10^{3} \text{K}$. The average $\text{H}_{2}$ density in the ``\textit{ray}'' CGM is higher by about 2 orders of magnitude compared to the ``\textit{apx-c}'' counterpart. Therefore, the use of the self-shielding model affects the temperature and density profiles of both the ISM and the CGM.  

\subsection{Comparison of Radially-Dependent Variables}
\label{subsect:radialprofile}

\begin{figure}
	\centering
	\includegraphics[width=0.95\columnwidth]{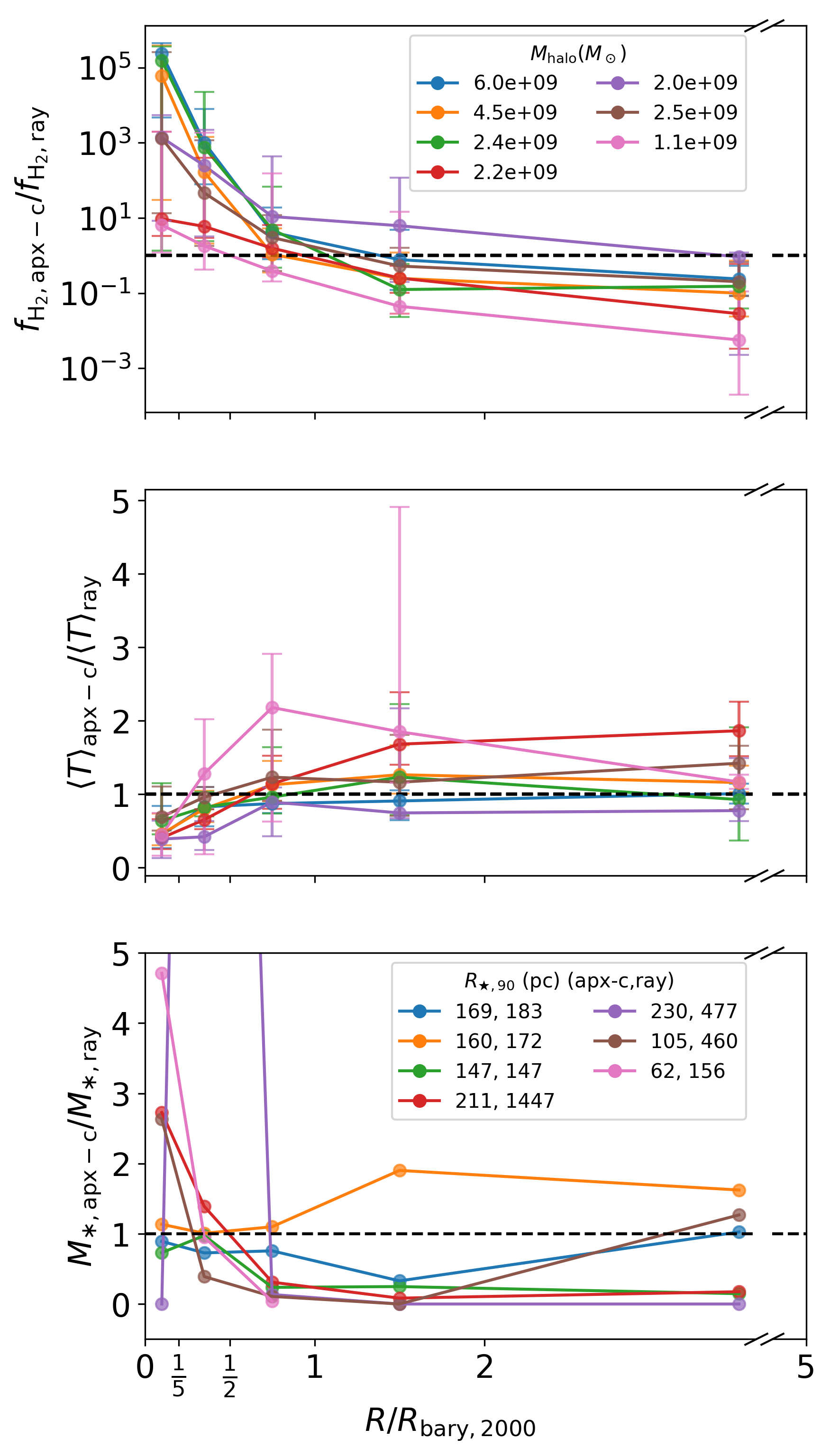}
	\caption{The comparison of the $\text{H}_{2}$ fraction ($f_{\text{H}_{2}}$), gas-mass-weighted average temperature ($\langle T \rangle$), and stellar mass in different radial bins between the ``\textit{apx-c}'' and the ``\textit{ray}'' simulations. For the $f_{\text{H}_{2}}$ and the gas temperature plot, the data points show the median of all ratio values since a halo starts to have stars, and the error bar shows the 25th to 75th percentile of those values. For the stellar mass plot, because the timing of star formation is slightly different between each simulation, we only evaluate at the last time step ($z = 12$) where the total stellar mass is converged within 0.5 dex between ``\textit{apx-c}'' and ``\textit{ray}''. Seven main progenitor halos (excluding sub-halos) are included. The $M_{200c}$ dark matter mass of each halo at $z = 12$ is listed in the legend of the top plot. The radius containing 90\% of the stellar mass ($R_{\bigstar,90}$) of each halo in both ``\textit{apx-c}'' and ``\textit{ray}'' is listed in the legend of the bottom plot. The approximation treatment affects the halo properties differently in different radial bins.}
	\label{fig:radialprofile}
\end{figure}

\begin{figure}
	\centering
	\includegraphics[width=1\columnwidth]{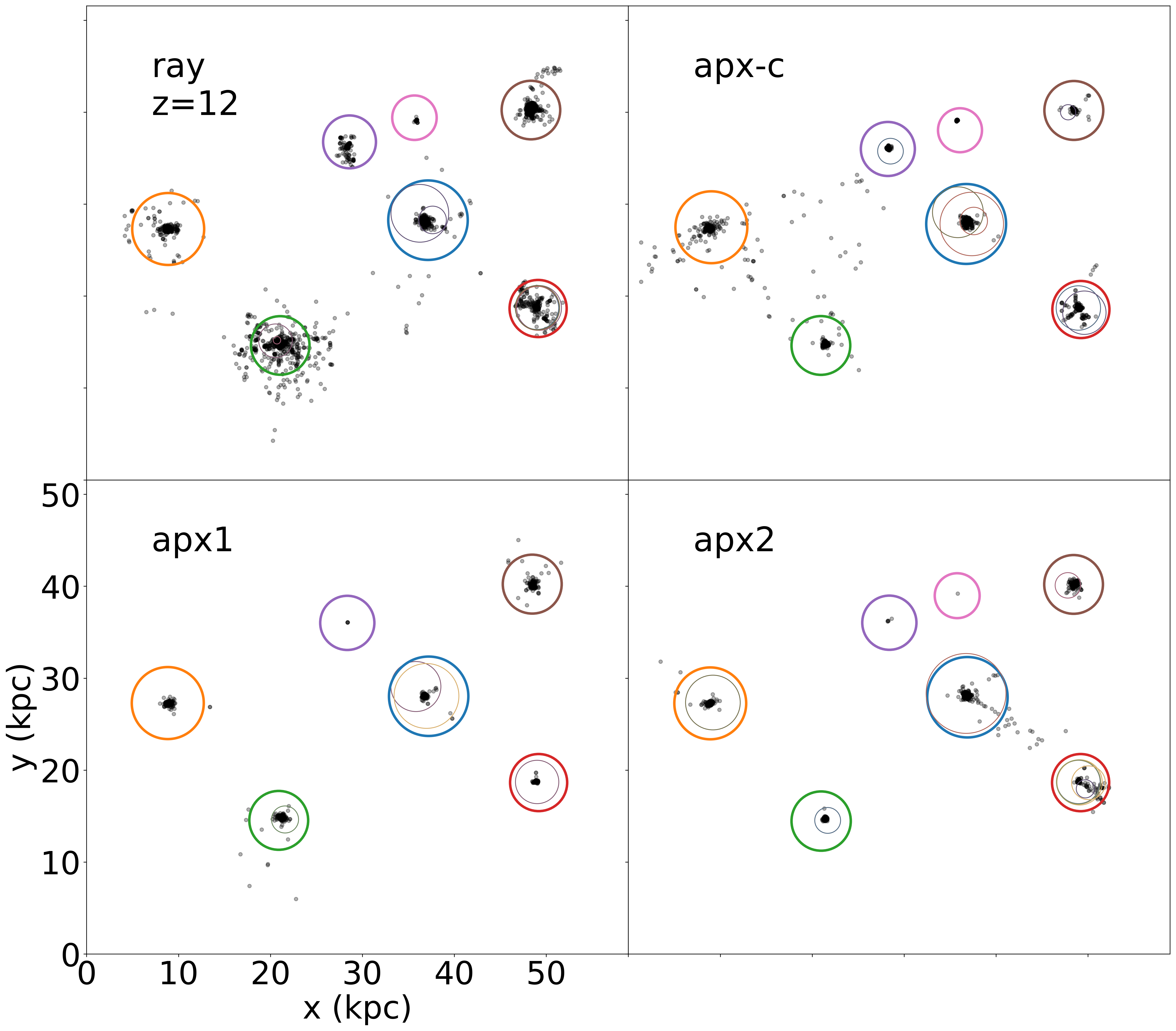}
	\caption{The positions of all star particles at $z = 12$ between our simulations. All halos that have stars bound to them are shown by circles. The thick circles represent main, isolated halos; the thin circles represent sub-halos. Cross-matched halos between the four simulations are colored similarly to each other and to the colors in the legends of Fig.~\ref{fig:radialprofile}. Among the seven cross-matched halos, the ``\textit{ray}'' halos tend to have a more extended distribution of star particles.}
	\label{fig:starposition_allbox}
\end{figure}

The effect of the self-shielding model on different regions of the halo is further illustrated in Fig.~\ref{fig:radialprofile}, which shows a comparison between ``\textit{apx-c}'' and ``\textit{ray}'' with respect to the $\text{H}_{2}$ fraction ($f_{\text{H}_{2}}$), gas temperature, and the stellar mass in five radial bins ranging from the center to approximately the radius $R_{200c}$ of each halo ($R_\text{200c}\approx 5R_\text{2000,bary}$). By $z = 12$, our simulations have seven main isolated halos (i.e., they are not subhalos to other bigger halos) that host stars, and all of them are included in this comparison. Because $\text{H}_{2}$ and gas temperature are sensitive to stellar feedback, and the timing of stellar feedback can be different between the simulations, for each radial bin, we calculate the $f_{\text{H}_{2}}$ ratio and the gas temperature ratio between ``\textit{apx-c}'' and ``\textit{ray}'' for all timesteps after the first stars form in a given halo. Then, we plot the median of all those ratios, and the error bars represent the 25th percentile to the 75th percentile of those ratios. This ensures that any instantaneously large difference in the ratio due to stellar feedback can be excluded, while we still capture the overall trend of the radial differences. This figure quantifies our qualitative analyses of Figs.~\ref{fig:Halo0-0_comparison} and \ref{fig:phaseplot_Halo0-0} and emphasizes that the Sobolev-like approximation affects the inner part and the outer part of a halo differently. The top subplot of Fig.~\ref{fig:radialprofile} demonstrates that the ratio of $f_{\text{H}_{2}}$ between the approximation and the ray-tracing model shows a clear inverse relationship with radius. For all seven halos, the ``\textit{apx-c}'' version predicts the $f_{\text{H}_{2}}$ of the most inner part of the galaxy to be 1 to 5 orders of magnitude higher than that of the ``\textit{ray}'' version. Also, the larger the halo is, the higher the gas density in the inner region is, resulting in a larger overestimation. As we go out to the CGM ($R_{\mathrm{bary}, 2000} < R < 5R_{\mathrm{bary}, 2000}$), gas density decreases and thus the approximation stops over-shielding $\text{H}_{2}$, leading to a decrease in the ratio of $\text{H}_{2}$ fraction. The $\text{H}_{2}$ fraction matches the best between the two models from 0.5$R_\text{2000,bary}$ to 2$R_\text{2000,bary}$. Further into the CGM, the approximation model significantly \corr{lowers} the degree of self-shielding, leading to an absence of $\text{H}_{2}$ in this region when compared with the ray-tracing model. This underestimation leads to a 1 to 3 order of magnitude difference in the CGM's $\text{H}_{2}$ fraction. 

The radial difference in $\text{H}_{2}$ between the models contributes to a radial difference in the temperature profile of the halos as well. Because more $\text{H}_{2}$ means that gas at $10^{4} \text{K}$ can cool down to a lower temperature, the mass-weighted average temperature at the inner part of the galaxy is also about half as high in the approximation model as in the ray-tracing model. At outer radii, the temperature converges better between the two models as the ratio $\langle{T}\rangle_\mathrm{apx-c}/\langle{T}\rangle_\mathrm{ray}$ approaches 1. At the outskirts of the halos, because we have less $\text{H}_{2}$, we see that the gas temperature in the approximation model is slightly higher than in the ray-tracing model. 

Regarding the radial distribution of stellar mass, we plot the ratio of the stellar mass in each radial shell at $z = 12$ in the bottom row of Fig.~\ref{fig:radialprofile}. Although the total stellar mass mostly agrees between simulations (Fig.~\ref{fig:stars}), we notice a mismatch in the radial distribution. Four out of seven halos in our analysis demonstrate a higher concentration of stars within the halo's $0.2R_\mathrm{bary,2000}$ and a lower amount of stars beyond $0.5R_\mathrm{bary,2000}$ in the ``\textit{apx-c}'' than in the ``\textit{ray}'' simulation. In other words, galaxies in the ``\textit{ray}'' simulation tend to be more extended, whereas galaxies in the \textit{apx} simulations tend to be more compact. This effect can be seen more clearly in Fig.~\ref{fig:starposition_allbox}, where we plot the position of all star particles in our volume and the host halos. The colors of the seven main halos match the line colors used in Fig.~\ref{fig:radialprofile}. Fig.~\ref{fig:starposition_allbox} displays that stars in galaxies with the ray-tracing model are more likely to be distributed widely into the CGM and even the IGM. In fact, at $z = 12$, about 2.4\% of stars are unassigned to any of the halos in the ``\textit{ray}'' simulations, meaning that they are beyond $R_{200c}$ of any halos and become unbound. On the other hand, the percentages of unassigned stars are significantly smaller, about 0.74\%, 0.18\%, and 0.40\% for ``\textit{apx-c}'', ``\textit{apx1}'', and ``\textit{apx2}'', respectively. Furthermore, since their stellar mass relatively matches, the ``\textit{ray} galaxies'' will be more diffuse than their \textit{apx} counterparts. Indeed, the 90\%-stellar-mass radius ($R_{\bigstar,90}$) is always higher by up to 7 times for halos in the ``\textit{ray}'' simulation, as displayed in the legends of the bottom subplot of Fig.~\ref{fig:radialprofile}. This observation matches with our expectation since the ``\textit{apx-c}'' run has more $\text{H}_{2}$ and lower temperature at the galaxy's inner core, making it easier for stars to form. The trend becomes opposite at an outer radius because the ``\textit{ray}'' run has more $\text{H}_{2}$ in this region. It is important to note that, up to $z = 12$, the overdensity threshold for star formation remains within an accuracy range of the approximation model (Subsection~\ref{subsect:effect_on_sf}). Therefore, the difference in stellar mass radial distribution is not as strong as in the $\text{H}_{2}$ fraction distribution.

\subsection{Effects on the Larger Volume and Reionization}
\label{subsect:larger_volume}

\begin{figure*}
	\centering
	\includegraphics[width=0.93\textwidth]{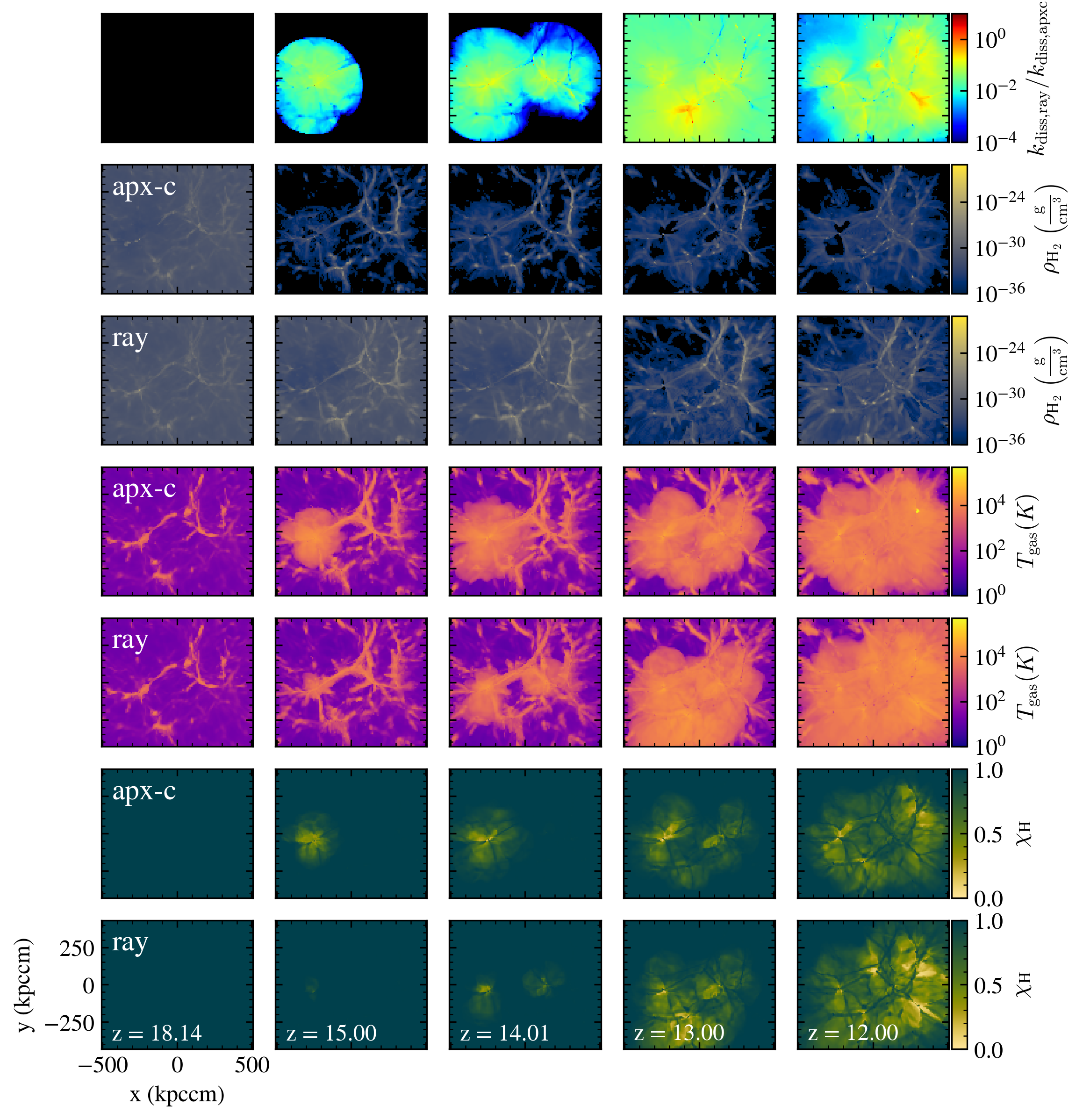}
	\caption{The ratio between the photodissociation rate computed by the ray-tracing method and the rate approximated by the Sobolev-like method (1st row). $\text{H}_{2}$ density (2nd and 3rd row), gas temperature (4th and 5th row), and the neutral hydrogen fraction (6th and 7th row) within the 5-$R_{200c}$ region between the ``\textit{apx-c}'' and ``\textit{ray}'' at different redshifts.}
	\label{fig:gas_surface_density_wholebox}
\end{figure*}

\begin{figure*}
	\centering
	\includegraphics[width=0.95\textwidth]{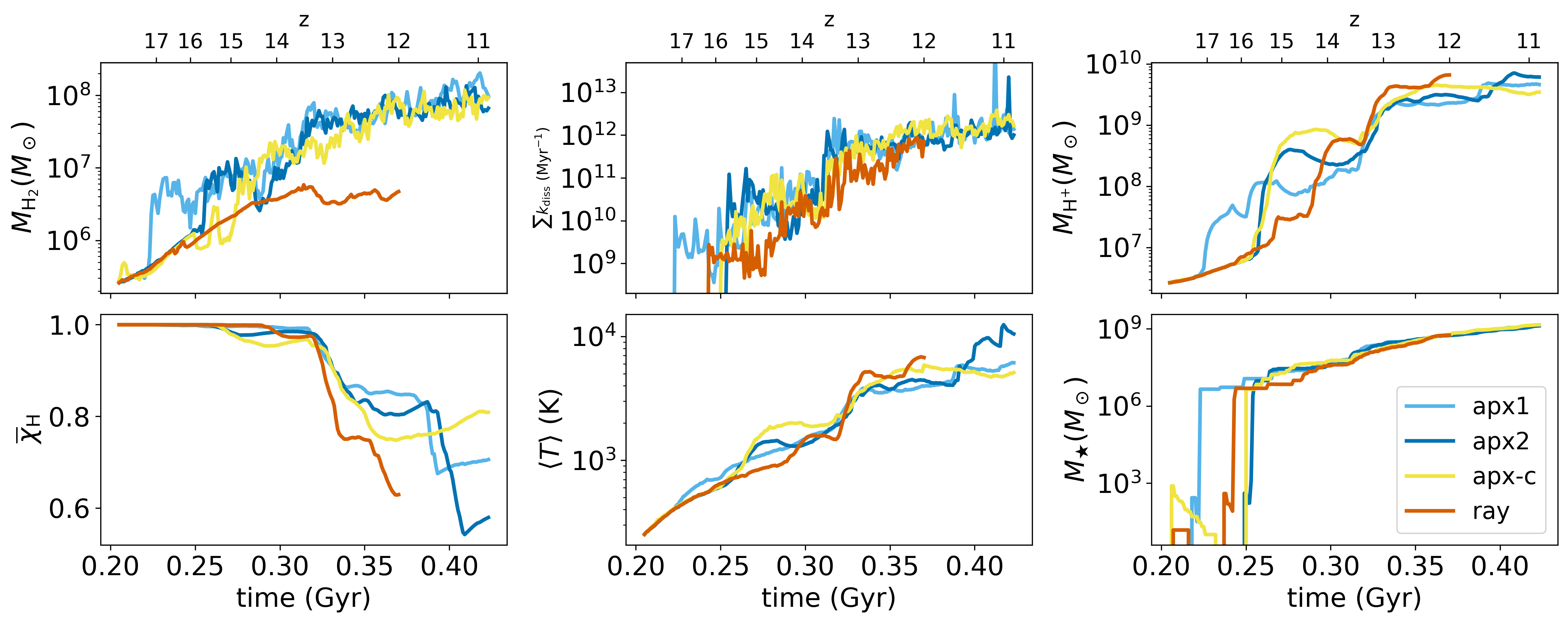}
	\caption{The comparison of different properties within the 5-$R_{200c}$ region as a function of time. From left to right, then top to bottom, the properties are neutral molecular hydrogen \corr{$\text{H}_{2}$}, total photodissociation rate $\sum k_\mathrm{kdiss}$, ionized atomic hydrogen \corr{$\text{H}^{+}$}, volume-weighted neutral hydrogen fraction \corr{$\overline\chi_{\mathrm{H}}$}, mass-weighted average gas temperature $\langle T \rangle$, and total stellar mass $M_\ast$.}
	\label{fig:wholebox_evolution}
\end{figure*}

\begin{figure*}
	\centering
	\includegraphics[width=0.99\textwidth]{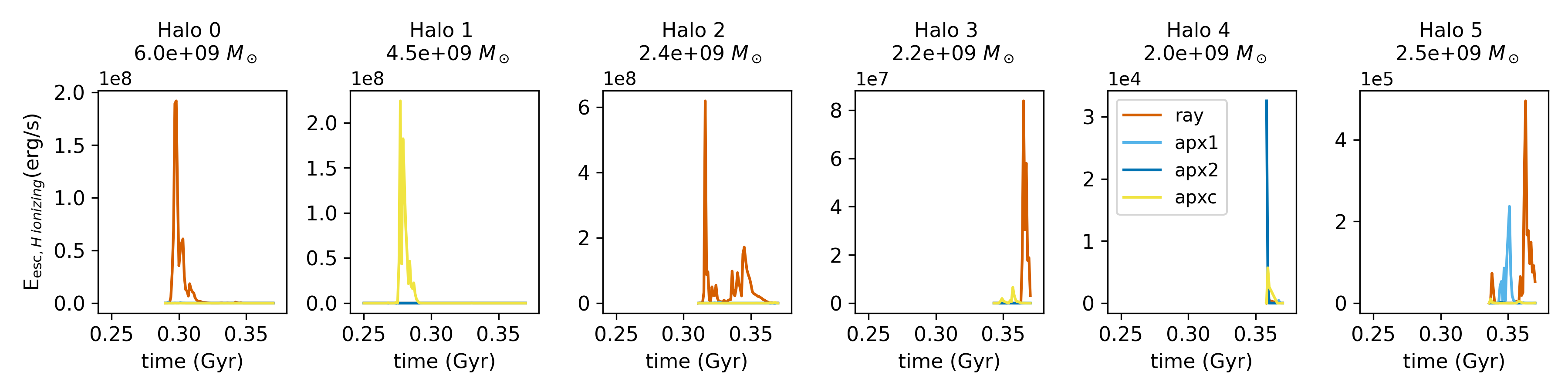}
	\caption{The escaping energy of \corr{H} ionizing radiation of six out of seven main progenitor halos in our simulation as a function of time. The last halo is omitted from the plot because there is no \corr{H} ionizing radiation escaping out of the halo in all four simulations. Among our four simulations, the ``\textit{ray}'' simulation has the most ionizing radiation escaping to the IGM.}
	\label{fig:escape_fraction_evolution}
\end{figure*}

We extended our analysis to the whole simulation box to investigate the effect of $\text{H}_{2}$ self-shielding modeling on a larger volume. Because our simulations are zoom-in simulations, stars only form inside the refined region, and only halos existing inside the refined region are considered physical because they are modeled with the best resolution. Also, gas cells and dark matter particles near the refined region's boundary typically have a lower resolution than the ones in the center, which makes star formation less likely. Thus, we limited our analysis to the region where stars exist inside the box. Within the zoom-in refined region, we defined a rectangular sub-volume that bounds the halos with stars and their local environments out to five times their $R_{200c}$.
%This region is determined as follows. First, for each halo with stars, we determine the upper right and lower left coordinates of a cube circumscribing the halo's 5-$R_{200c}$ sphere. Then, we determine the upper right and lower left coordinates of the analyzed region by finding the maximum of all cubes' upper right coordinates and the minimum of all cubes' lower left coordinates. This region hence covers all halos with stars up to five times of their virial radii, and we check to make sure that the region still rests within the simulation's refined region. 
This region is about 1/6500 volume of the whole simulation box. For convenience, we will refer to this region as the 5-$
R_{200c}$ region. Fig.~\ref{fig:gas_surface_density_wholebox} shows the projection plots at different snapshots of the 5-$R_{200c}$ region. It is important to note that zoom-in simulations do not show the average properties of the universe but rather only the effects inside the zoomed-in region, which is prone to the properties of the halo population within that region. 

While we see a relatively similar filament structure between the ``\textit{ray}'' and ``\textit{apx-c}'' simulations, the distribution of $\text{H}_{2}$ and gas temperature does differ. As shown in the second and third rows of Fig.~\ref{fig:gas_surface_density_wholebox}, the amount of $\text{H}_{2}$ in the IGM is much lower when using the approximation after stars start to form. Newly formed stars produce LW radiation that propagate the whole simulation's refined region and destroy $\text{H}_{2}$ in the low gas density region where the approximation model under-predicts the amount of shielding. The first row of the plot displays the ratio between the photodissociation rate obtained in the ``\textit{ray}'' simulation and the photodissociation rate manually calculated for those same gas cells if we use the Sobolev-like approximation (similar to the procedure in Fig.~\ref{fig:kdiss_apxc-ray_comparison}). The black color in the first-row plots of Fig.~\ref{fig:gas_surface_density_wholebox} represents regions where $k_{\text{diss,ray}}=0$. We see that the two rates match better inside the halos or along the filaments where the gas density is relatively high. In the IGM, $k_\text{diss,ray}$ is lower than $k_\text{diss,apx-c}$ by several orders of magnitude, which explains the discrepancy between the two $\text{H}_{2}$ projection plots (second and third rows of Fig.~\ref{fig:gas_surface_density_wholebox}). Furthermore, it is not until after $z = 14$ that the whole 5-$R_{200c}$ region has a non-zero photodissociation rate in the ``\textit{ray}'' simulation. When we use the approximation model, because we assume an optically thin LW radiation field and assume no light travel time delay, given that the zoom-in volume is relatively small, the whole region will have a non-zero photodissociation rate right after the first star appears. On the other hand, the ``\textit{ray}'' simulation traces rays through dense $\text{H}_{2}$ regions, and those regions can attenuate all LW photons before they reach further gas cells. This explains why the ``\textit{apx-c}'' simulation loses so much $\text{H}_{2}$ in the ISM early in the run ($z \geq 15$) while the ``\textit{ray}'' simulation does not. 

As shown in the top left subplot in Fig.~\ref{fig:wholebox_evolution}, the ``\textit{ray}'' simulation has a significantly lower amount of $\text{H}_{2}$ than the ones with the approximation treatment, especially after stars are created. In the top middle subplot, we also show the total photodissociation rate within the 5-$R_{200c}$ region obtained individually from ENZO for each simulation (instead of manually calculated on the ``\textit{ray}'' simulation like in Fig.~\ref{fig:kdiss_apxc-ray_comparison} and \ref{fig:gas_surface_density_wholebox}). The result shows that the ``\textit{ray}'' simulation experiences a lower photodissociation rate overall because most of the simulation volume is covered in low-density and moderate-density gas ($\rho_\text{gas} \leq 10^{-20} \text{g}/\text{cm}^{3}$) and the approximation \corr{computes a lower} $k_\text{diss}$ in those areas. Despite a lower total photodissociation rate, the ``\textit{ray}'' simulation still has a deficit in molecular gas because in dense gas regions where $\text{H}_{2}$ exists the most, the ray-tracing model has a higher photodissociation rate. Similar to Subplots (b) of Fig.~\ref{fig:Halo0-0_comparison}, the difference in $\text{H}_{2}$ mass between the two radiative models is then compensated by an increase of \corr{$\text{H}^{+}$}. The stellar mass in the whole simulation box is slightly lower when using the ray-tracing models, but generally converges between the four simulations, as was the case in our analysis in Subsection~\ref{subsect:effect_on_sf}. 

Another behavior we can see is that after being destroyed by supernova feedback, $\text{H}_{2}$ recombines in the supernova-induced \corr{$\text{H}^{+}$} region and in the galactic filaments. In the temperature projection plot (fourth and fifth rows of Fig.~\ref{fig:gas_surface_density_wholebox}), supernova explosions heat up and ionize the surrounding gas. $\text{H}_{2}$ can then be destroyed through thermal dissociation or photodissociation. On the other hand, the electrons and ionized hydrogen created from the ionization provide ingredients for $\text{H}_{2}$ to form (see equations in the chemical network description in Subsection~\ref{subsec:cosmological_simulations}). These explain the temporary absence of $\text{H}_{2}$ close to galaxies after supernova feedback and the accumulation of $\text{H}_{2}$ around galaxies and galactic filaments in Fig.~\ref{fig:gas_surface_density_wholebox}. 
%At high gas density region inside galaxies, $\text{H}_{2}$ can form via a three-body interaction between three atomic hydrogen \citep{Palla+1983}. %\textbf{consider deleting this sentence}

Regarding the mass-weighted mean gas temperature (bottom middle plot of Fig.~\ref{fig:wholebox_evolution}), we do not see a considerable difference between the two radiative treatments within the 5-$R_{200c}$ region. No simulation is always hotter than others, and thus any differences that we see are likely due to stellar feedback going off at different times in different simulations. Even though at a small scale, the higher amount of $\text{H}_{2}$ in the galactic center can help cool off the gas in the approximation model, the region is too small to affect the overall mean temperature. This additional cooling by more $\text{H}_{2}$ in the galactic center can be balanced off by the lower amount of $\text{H}_{2}$ in the IGM in the \textit{apx} simulations. 

We also investigated whether the use of the approximation model affects cosmic Reionization. The bottom two rows of Fig.~\ref{fig:gas_surface_density_wholebox} show the distribution of neutral hydrogen fraction \corr{$\chi_{\mathrm{H}}$} within the 5-$R_{200c}$ region between ``\textit{apx-c}'' and ``\textit{ray}'' simulations. The \corr{$\chi_{\mathrm{H}}$} value of 0 corresponds to complete ionization of \corr{H}. High-energy radiation from galaxies ionizes hydrogen, decreasing \corr{$\chi_{\mathrm{H}}$} in the surrounding IGM regions. \corr{$\chi_{\mathrm{H}}$} can go up again during recombination between protons and electrons to form neutral hydrogen, which is efficient at a temperature lower than $10^{4} \mathrm{K}$ \citep{Wise+2019}. Even though ``\textit{apx-c}'' has a slightly bigger region of ionized hydrogen in the beginning (second and third columns of the bottom two rows of Fig.~\ref{fig:gas_surface_density_wholebox}), at $z = 12$, the ``\textit{ray}'' simulation displays a more extended and prominent region of ionized hydrogen, expanding to encompass more of the IGM. This observation is quantified in the lower left subplot of Fig.~\ref{fig:wholebox_evolution}, where we plot the volume-weighted average of the neutral hydrogen fraction (\corr{$\overline{\chi}_{\mathrm{H}}$}) as a function of time. In the beginning, the universe is dominated by neutral hydrogen (\corr{$\overline{\chi}_{\mathrm{H}}$} $\approx 1$). However, about 0.26 Gyr after the Big Bang ($z \approx 15.4$), the neutral hydrogen fraction starts to drop. The \corr{$\overline{\chi}_{\mathrm{H}}$} of the \textit{apx} simulations drops first, but then ionized hydrogen gets recombined and \corr{$\overline{\chi}_{\mathrm{H}}$} increases or plateaus. In contrast, the \corr{$\overline{\chi}_{\mathrm{H}}$} of the ``\textit{ray}'' simulation declines very quickly after 0.3 Gyr after the Big Bang. By $z = 12$, 63\% of the atomic hydrogen in the 5-$R_{200c}$ region gets ionized in ``\textit{ray}'' while that value is 85\%, 80\%, 75\% in ``\textit{apx1}'', ``\textit{apx2}'', and ``\textit{apx-c}'', respectively. We again note that the reason why the result of ``\textit{apx1}'' and ``\textit{apx2}'' differ from each other is that our PopIII stellar mass relies on random sampling \citep{Bryan+2014}. Even if the gas profile is the same, each halo in ``\textit{apx1}'' and ``\textit{apx2}'' can form different PopIII mass, resulting in a different amount of supernova feedback and consequently a different amount of hydrogen being ionized. However, this discrepancy within the \textit{apx} simulations is still considerably smaller than the discrepancy between the ``\textit{ray}'' simulation and the \textit{apx} simulations, which is evidence that the ray-tracing method has a significant effect on Reionization. To explore more carefully why \corr{$\overline{\chi}_{\mathrm{H}}$} mismatches between the models, we calculate the escaping energy of the ionizing radiation of each isolated main halo, as shown in Fig.~\ref{fig:escape_fraction_evolution}. The method for computing the escaping ionizing energy is as follows. We first used the HEALPix for the \texttt{python} package \texttt{healpy} to trace unit vectors from the center of the halo to its boundary. To get the optical depth $\tau$ along each of these paths, we traced a \texttt{yt} ray originating from every star particle to the halo boundary. Then, we multiplied the star particle's intrinsic spectra by $\text{e}^{-\tau}$ to get the attenuated spectra, and integrated the spectra over the ionising frequencies to get the escaping energy. Intrinsic stellar spectra were calculated using Flexible Stellar Population Synthesis \citep{Conroy+2009, Conroy+2010} with a Davé IMF \citep{Dave+2008} for metal enriched (Population II) and YGGDRASIL \citep{Zackrisson+2011} for metal poor (Population III) stars. We use 55$M_\odot$ as a cutoff between the PopIII.1 and PopIII.2 (top-heavy) models.

In \corr{four} out of six halos with non-zero escaping ionizing energy, the ``\textit{ray}'' halos allow much more ionizing radiation to escape from the halos. Among the other \corr{two} halos, Halo 4 \corr{has} more ionizing radiation coming out in the ``\textit{apx2}'' and ``\textit{apx-c}'' simulations; however, these escaping energies are still much smaller than the amount of Halo 2 and Halo 3 of the ``\textit{ray}'' simulation emit at the similar time. Therefore, Fig.~\ref{fig:escape_fraction_evolution} suggests that the atomic hydrogen in the ISM is more likely to be ionized when using the ray-tracing models. This is also consistent with the fact that galaxies in the ``\textit{ray}'' simulation have a wider stellar distribution (shown in Fig.~\ref{fig:starposition_allbox}), hence gas in the outer CGM or ISM can be ionized more easily by these outer-radius stars. Another potential contribution to explain why \corr{$\overline{\chi}_{\mathrm{H}}$} is lowest in the ``\textit{ray}'' simulation is the conversion of $\text{H}_{2}$ to \corr{$\text{H}^{+}$} within the halo. Because the $\text{H}_{2}$ near the halo center in ``\textit{ray}'' simulation has less self-shielding, it gets photodissociated more easily into \corr{H} and then \corr{$\text{H}^{+}$}, hence decreasing the neutral hydrogen fraction. While the simulations are not run long enough to cover the whole Reionization epoch, the bottom left subplot of Fig.~\ref{fig:wholebox_evolution} suggests that reionization of our 5-$R_{200c}$ region may be complete earlier in the simulation with ray-tracing than in the simulations with the approximation model, even though the Reionization's start time is relatively similar. \corr{However, as pointed out in the case of star formation in Section~\ref{subsect:effect_on_sf}, as gas metallicity increases and gets distributed more sidely at later times in galaxies, this may alleviate the dependence of star formation on the $\text{H}_{2}$ photodissociation rate. As a result, this reduces the differences in stellar mass and the radial stellar distribution between galaxies in our two $\text{H}_{2}$ self-shielding models, and the $\overline{\chi}_{\mathrm{H}}$ value of our simulations may converge back at lower redshifts.} 

%This illustrates that the use of Sobolev-like approximation to model $\text{H}_{2}$ self-shielding may slow down the reionization process and thus affects our theoretical prediction and modeling of the Reionization epoch. 

\section{Comparisons with Previous Works}
\label{sec:comparison}
 
Prior works examined the application of the Sobolev-like density-gradient approximation on a more local scale with varying results. When simulating dark matter halos with an LW background, \cite{Wolcott-Green+2011} showed that the Sobolev-like approximation can underestimate the self-shielding factor $f_\text{sh}$ in the high-$f_\text{sh}$ regime and slightly overestimate $f_\text{sh}$ in the low-$f_\text{sh}$ regime. This translates to a lower approximated $k_\text{diss}$ in the high-$k_\text{diss}$ regime and a slightly higher approximated $k_\text{diss}$ in the low-$k_\text{diss}$ regime when using the approximation. This generally matches with the behavior in the top left subplot of our Fig.~\ref{fig:kdiss_apxc-ray_comparison}. However, because the authors assume isotropy of constant LW intensity background on their halos, their dense gas cells always have a lower photodissociation rate than their diffuse gas cells. Therefore, since our LW field is anisotropic with stars as radiative sources, our result concerning the relationship between the photodissociation rate and gas density is not comparable with theirs. \cite{Safranek-Shrader+2017} investigated different $\text{H}_{2}$ radiative transfer models with an idealized galactic disc simulation and found that compared to the ray-tracing, the Sobolev-like approximation results in a higher effective extinction \corr{in the V band} at high density ($n > 5\times10^{2}\,\text{cm}^{-3}$) and lower extinction at low density ($n < 10^{2}\,\text{cm}^{-3}$). Because the $v$-band extinction is proportional to the sum of the column density of \corr{H} and $\text{H}_{2}$ in their study, this implies that there are more $\text{H}_{2}$ at high gas density and less $\text{H}_{2}$ at low gas density. The number density values in their result correspond to the gas density of $\approx 10^{-22} - 10^{-21} \text{g}/\text{cm}^{3}$, which matches with the gas density threshold where the approximation starts to \corr{boost} $\text{H}_{2}$ mass is our analysis (top middle subplot of Fig.~\ref{fig:kdiss_apxc-ray_comparison}). 

On the other hand, through studying the formation of PopIII stars in the D-type ionization front surrounding a star in a dark matter minihalo, \cite{Chiaki+2023a} found that the density gradient approximation matches well with the ray-tracing scheme. However, given the limited physical scale of their phenomenon ($\approx 100\;\text{pc}$), the limited radiative source (only one primary star), and the limited time scale ($t < 0.1$ Myr), their findings are not directly comparable to our scales.

\section{\corr{Caveats}}

\corr{We identify three caveats of this work. Firstly, our mass resolution is not high enough to resolve cold gas in mini halos with mass $\leq 2\times10^{6} M_\odot$, even though halos in this mass range are still able to host stars \citep{CorreaMagnus+2024}. Although we expect that the Sobolev-like approximation will inhibit star formation in mini halos, we cannot fully explore the effects of the self-shielding models for these halos.} 

\begin{figure}
	\centering
	\includegraphics[width=1\columnwidth]{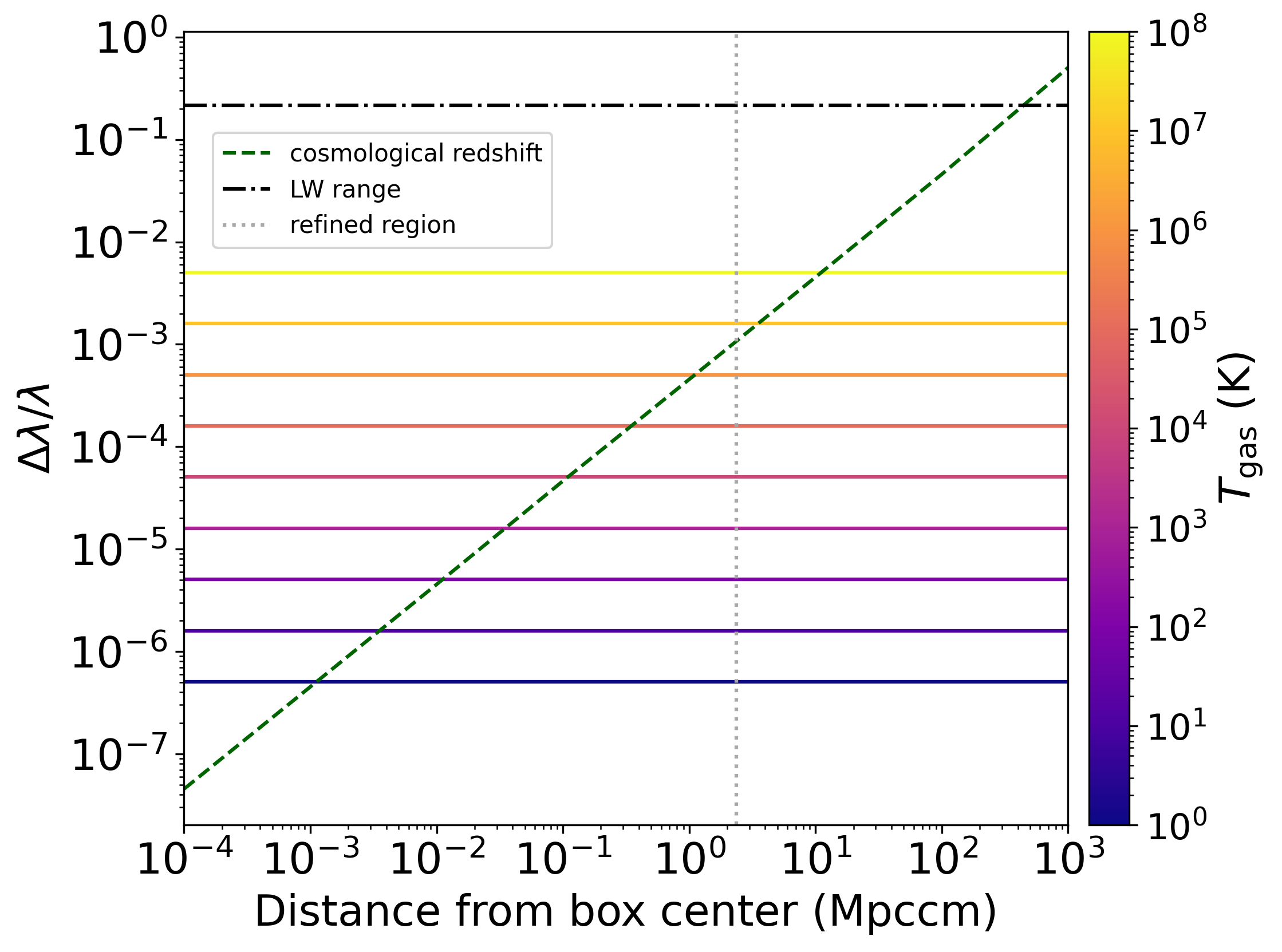}
	\caption{\corr{A comparison between the FWHM thermal broadening of $\text{H}_{2}$ molecules at different temperatures (solid lines) and the cosmological redshift as viewed from the box center due to the Hubble expansion (dashed green lines). The bottom x-axis shows the distance in comoving units. The vertical dotted lines represent the refined region's boundary. The horizontal dash-dotted black line represents the redshift at which a rest-frame LW photon is shifted out of the LW band. Within the refined region, even though most of the gas cells have the thermal velocity smaller than the recession velocity, the ray-tracing modeling is still not significantly affected by the cosmological redshift of the Hubble expansion because of the large range of the LW band.}}
	\label{fig:thermal_vs_doppler_velocity}
\end{figure}

\corr{Secondly, even though the adaptive ray-tracing model is more sophisticated than the Sobolev-like approximation and includes the thermal broadening effect, it currently does not account for the Doppler shifting effect of the Hubble expansion. Hubble expansion results in a velocity difference between $\text{H}_{2}$ molecules in different regions in the simulation box, causing the photon energy to be shifted out of the $\text{H}_{2}$'s absorption band and decreasing the effective $\text{H}_{2}$ column density for self-shielding. In other words, by neglecting this effect, the ray-tracing model may overestimate the actual amount of \corrtwo{the incident LW flux}. To examine this caveat in our work, we show the comparison between the full width at half maximum (FWHM) of the thermal broadening and redshift due to Hubble expansion as viewed from the box center in Fig.~\ref{fig:thermal_vs_doppler_velocity}. The solid lines show thermal broadening of $\text{H}_{2}$'s absorption lines at different temperatures. The dashed lines show the Hubble expansion's redshift of the photon as it travels from the box center to a certain distance away. The figure shows that the Hubble expansion recessional velocity dominates over the thermal velocity of the gas, especially for cold gas cells that are distant from the radiative source. Nevertheless, the redshift caused by the Hubble expansion is still relatively small, which is only up to $\approx 0.001$ at the refined region's boundary. Given a wide range of the LW band ($\approx 91\text{--}111$ nm), the redshift needs to be $\approx 0.22$ (shown by the dash-dotted black line in the plot) in order for a photon to be shifted out of the range. In other words, at the refined region of our simulations, only about 0.5\% of our photons \corrtwo{fall} out of the LW range. \corrtwo{The distribution of $\text{H}_{2}$ between a target gas cell and the radiative sources is another key factor that can influence the accuracy of the ray-tracing scheme. If stellar LW photons are heavily redshifted before reaching a distant collection of $\text{H}_2$, the ray-tracing method would tend to overestimate the dissociation rate. However, as shown in Fig.~\ref{fig:gas_surface_density_wholebox}, most $\text{H}_{2}$ in our simulations is distributed near the box center as well as near the radiative sources. Fig.~\ref{fig:kdiss_apxc-ray_comparison} also emphasizes that the photodissociation rate is higher at denser gas cells located nearer the radiative sources due to higher fluxes received by more local stellar sources. This means that most of the photodissociation and self-shielding effect happens locally and that very little of the $\text{H}_{2}$ providing shielding is considerably redshifted relative to the target gas cells further away.}
%Moreover, the boundary of the 5-$R_{200c}$ region in Section~\ref{subsect:larger_volume} is only about half the distance to the refined region's boundary, showing that the ray-tracing model for $\text{H}_{2}$ is still valid against the Hubble expansion in this region. 
\corrtwo{Lastly, another} reason why this caveat can be minor is that for the ENZO ray tracing scheme, if a photon travels a long distance and the photon flux falls below a background radiation value, it will be deleted. Therefore, in \corrtwo{the refined region of} our simulation sets, the ray-tracing method is likely to provide a more accurate measure of the $\text{H}_{2}$ self-shielding than the Sobolev-like approximation. However, for simulations with the same redshift range whose size is larger than 450 Mpccm (or about 35 Mpc in proper unit at $z = 12$), \corrtwo{when $\text{H}_{2}$ is more populated between stars and distant gas cells}, and when the flux is significant, it may be important to consider the cosmological redshift, and caution is needed when using the ray-tracing method for $\text{H}_{2}$ self-shielding.}

\corr{As remarked in Subsections~\ref{subsect:effect_on_sf} and ~\ref{subsect:larger_volume}, the third caveat is that we do not fully explore the contribution of cooling by metals and other heavier molecules in this work. At lower redshift, the ISM will be more enriched with metals, and these new cooling channels may dominate over $\text{H}_{2}$ cooling. Thus, the discrepancies that we identify in this work may become less visible, especially with regard to star formation. Future work is needed to explore this question.}

\section{Conclusion}
\label{sec:conclusion}

In this paper, using high-resolution zoom-in cosmological simulations, we re-simulate the same volume with two radiative transfer models of $\text{H}_{2}$ self-shielding to examine their effect on the simulated universe. The two models are the Sobolev-like desntiy-gradient approximation that uses local gas properties to estimate the $\text{H}_{2}$ photodissociation rate and the adaptive ray-tracing model. 
%We also corrected a bug in the previous ENZO implementation of the Sobolev-like approximation and examined whether the correction impacts the outcome. 
In summary, our key findings are as follows:

\begin{itemize}
	\item \corr{Compared to the ray-tracing method, the Sobolev-like approximation shows a lower amount of $\text{H}_{2}$ self-shielding at low gas density ($\rho_\mathrm{gas} \leq 10^{-22} \mathrm{g}/\mathrm{cm}^{3}$) while shows more $\text{H}_{2}$ self-shielding at high gas density ($\rho_\mathrm{gas} \geq 10^{-20} \mathrm{g}/\mathrm{cm}^{3}$)}. This leads to low-mass halos ($M_\text{halo} \leq 10^{8}M_\odot$) in the \textit{apx} simulations having their $\text{H}_{2}$ mass several orders of magnitude lower than their counterparts in the simulation with the ray-tracing model. On the other hand, high-mass halos ($M_\text{halo} \geq 10^{9}M_\odot$) have their $\text{H}_{2}$ mass several of magnitude higher in the approximation model than in the ray-tracing model (Figs.~\ref{fig:kdiss_apxc-ray_comparison} and \ref{fig:H2_frac}). 
        \item The correction of the ENZO implementation of the Sobolev-like approximation does not change the photodissociation rate significantly and does not seem to affect the global results. Previous runs with ENZO using the old implementation can still be representative of the Sobolev-like approximation. Also, different realizations of the same simulation parameters still converge in the halo properties and the amount of molecular gas in the box (Figs.~\ref{fig:kdiss_apxc-apx_comparison}, \ref{fig:Halo0-0_comparison}, and \ref{fig:wholebox_evolution}).
	\item The approximation inhibits star formation in small halos ($M_\text{halo} < 10^{9}M_\odot$). On the other hand, the stellar mass of more massive halos converges better between the two radiative models. (Fig.~\ref{fig:stars}).  
	\item Near the galactic center, the approximation results in denser and colder molecular clouds and a higher concentration of stellar mass. Out in the CGM, the galaxies in the approximation model have less $\text{H}_{2}$ and fewer stars. This makes galaxies run with the approximation model more compact (Figs.~\ref{fig:radialprofile} and \ref{fig:starposition_allbox}). Also, the ray-tracing method results in a higher percentage of stars in the IGM and unbound with respect to the $R_{200c}$ of any halos. 
	\item The Sobolev-like approximation \corr{may affect} ionization fraction timings in the Reionization epoch, where it reionizes the universe slower than the prediction using the ray-tracing model. (Fig.~\ref{fig:wholebox_evolution})
\end{itemize}

These results show that using the Sobolev-like approximation has a considerable impact on predictions made with radiative-hydrodynamic cosmological simulations, particularly with regard to $\text{H}_{2}$, star formation in small halos, galaxy compactness, and the Reionization period. Studies on the gas density-temperature profile of ISM and CGM will also likely be affected by using the approximation. Therefore, predictions based on this model of $\text{H}_{2}$ self-shielding should be re-evaluated as needed. This comparison could be expanded to other self-shielding approximation models and other formulas for the self-shielding factors, as discussed in Section~\ref{sec:intro}. It is possible that one method performs better in one aspect and worse in another, which makes it important for simulation users to be aware of and apply the appropriate method accordingly to their scientific questions.

%% Also note that the akcnowlodgment environment does not support long amounts of text. If you have a lot of people and institutions to acknowledge, do not use this command. Instead, create a new \section{Acknowledgments}.
\section*{ACKNOWLEDGEMENTS}
THN acknowledges support from the National Center for Supercomputing Applications Center and its Center for Astrophysical Surveys. THN, KSSB, VS and SB acknowledge the University of Illinois at Urbana-Champaign for their continued support. \corrtwo{We thank the referee for their proofreading and valuable suggestions to improve the work}.  We thank John Wise for the helpful discussion and response on the ENZO implementation of $\text{H}_{2}$ self-shielding, and we thank Gen Chiaki for sharing the code of their work on $\text{H}_{2}$ self-shielding publicly on GitHub. THN and KSSB acknowledge the support of the Delta Supercomputer as well as the ACCESS program for computing grants PHY230100 and PHYS240175.  

\section*{DATA AVAILABILITY}
The raw simulation data, the analysis code, and the simulation metadata underlying this project can be shared on reasonable request to the corresponding author.


\vspace{5mm}
%\facilities{HST(STIS), Swift(XRT and UVOT), AAVSO, CTIO:1.3m}


%\software{astropy, Cloudy, Source Extractor}




\bibliography{sample631}{}
\bibliographystyle{aasjournal}



\end{document}